\documentclass[aps,twocolumn,superscriptaddress,longbibliography,amsmath,amssymb,amsfonts,citeautoscript]{revtex4-1}
\usepackage{graphicx}
\usepackage{bm}
\usepackage{bbm}
\usepackage{color}
\usepackage{epstopdf}
\usepackage{amsmath}
\usepackage{amssymb}
\usepackage{ulem}
\usepackage[urlcolor=blue,colorlinks=true,citecolor=blue,linkcolor=blue,pdfstartview={FitH},bookmarks=false]{hyperref}
\usepackage{todonotes}
\usepackage{ulem}

\newcommand{\expect}[1]{\langle #1 \rangle}

\begin{document}
\title{Transient effects in quantum dots contacted via topological superconductor}

\author{R. Taranko}
\affiliation{Institute of Physics, M. Curie-Sk\l{}odowska University, 20-031 Lublin, Poland}

\author{K. Wrze\'sniewski}
\affiliation{Institute of Spintronics and Quantum Information, Faculty of Physics, A.~Mickiewicz University, 61-614 Pozna{\'n}, Poland}

\author{I. Weymann}
\affiliation{Institute of Spintronics and Quantum Information, Faculty of Physics, A.~Mickiewicz University, 61-614 Pozna{\'n}, Poland}

\author{T. Doma\'nski}
\affiliation{Institute of Physics, M. Curie-Sk\l{}odowska University, 20-031 Lublin, Poland}

\date{\today}

\begin{abstract}
We investigate gradual development of the quasiparticle states in two quantum dots attached to opposite sides of the topological superconducting nanowire, hosting the boundary modes. Specifically, we explore the non-equilibrium cross-correlations transmitted between these quantum dots via the zero-energy Majorana modes. Our analytical and numerical results reveal the nonlocal features observable in the transient behavior of electron pairing, which subsequently cease while the hybrid structure evolves towards its asymptotic steady-state configuration. We estimate duration of these temporary phenomena. Using non-perturbative scheme of the time-dependent numerical renormalization group technique we also analyze nonequilibrium signatures of the correlation effects competing with the proximity induced electron pairing. These dynamical processes could manifest themselves in braiding protocols imposed on the topological and/or conventional superconducting quantum bits, using superconducting hybrid nanostructures. 
\end{abstract}

\maketitle

\section{Introduction}
Majorana quasiparticles localized at the edges of one-dimensional chains \cite{Kitaev_2001},  
propagating along the boundaries of two-dimensional topological superconductors 
\cite{Read-2000} and confined on internal defects in $p$-wave bulk superconductors 
\cite{Volovik-2014} have recently been the topic of intensive studies (see e.g.\ 
\cite{Aguado.2017,Sato_2017,Lutchyn-2018,Kouwenhoven-2020,Flensberg-2021,Yazdani.etal_2023} and references cited therein). 
Such research interests are motivated by the exotic character of Majorana modes
and additionally stimulated by perspectives for constructing stable quantum bits (immune to 
external perturbations due to topological protection) and for performing quantum computations 
(by virtue of their non-Abelian character) \cite{Alicea-2016}. Majorana 
quasiparticles, realized in various platforms, including the minimalistic 
Kitaev chain consisting of two and three coupled quantum dots \cite{Kouwenhoven-2023,Bordin2024Feb}, do always 
emerge in pairs. To what extent, however, these spatially separated zero-energy modes are 
cross-correlated either statically (enabling teleportation phenomena 
\cite{Tewari_2008,Fu_2010,Li_2020,teleportation-2022}) or dynamically is still a matter 
of controversy. Some theoretical studies have predicted that 
Majorana modes might exhibit their dynamical interdependence in 
the shot-noise \cite{Liu-2015,Beenakker_2020,Perrin-2021,Smirnov_2023}.
Yet, any evidence for such nonlocal cross-correlations is missing.

Another convenient platform for exploring the unique features of Majorana modes 
are hybrid structures, where single or multiple quantum dots are side-attached 
to topological superconductors \cite{Kouwenhoven-2020}. The natural tendency of
Majorana modes to be harbored at the outskirts of low-dimensional superconductors
gives rise to their leakage onto the attached quantum dot(s) 
\cite{Flensberg-2011,Baranger-2011,Lopez-2013,Lutchyn-2014,Vernek-2014}. 
Such leakage has been indeed reported experimentally \cite{Deng-2016}, 
stimulating further theoretical studies concerning the interplay of Majorana
modes with the correlation-driven effects \cite{Vernek-2015,Prada-2017,Klinovaja-2017,
Ptok-2017,Weymann-2017,Deng-2018,Silva-2020,Seridonio-2020,Majek_Weyman-2021,
Domanski-2023,Vernek-2023,Cayao_2023}. Here we propose to consider similar
hybrid structures, comprising two quantum dots interconnected through the 
topological superconductor, in order to search for possible signatures of 
their nonlocal feedback effects appearing under non-equilibrium conditions.

\begin{figure}[b]
\includegraphics[width=0.95\columnwidth]{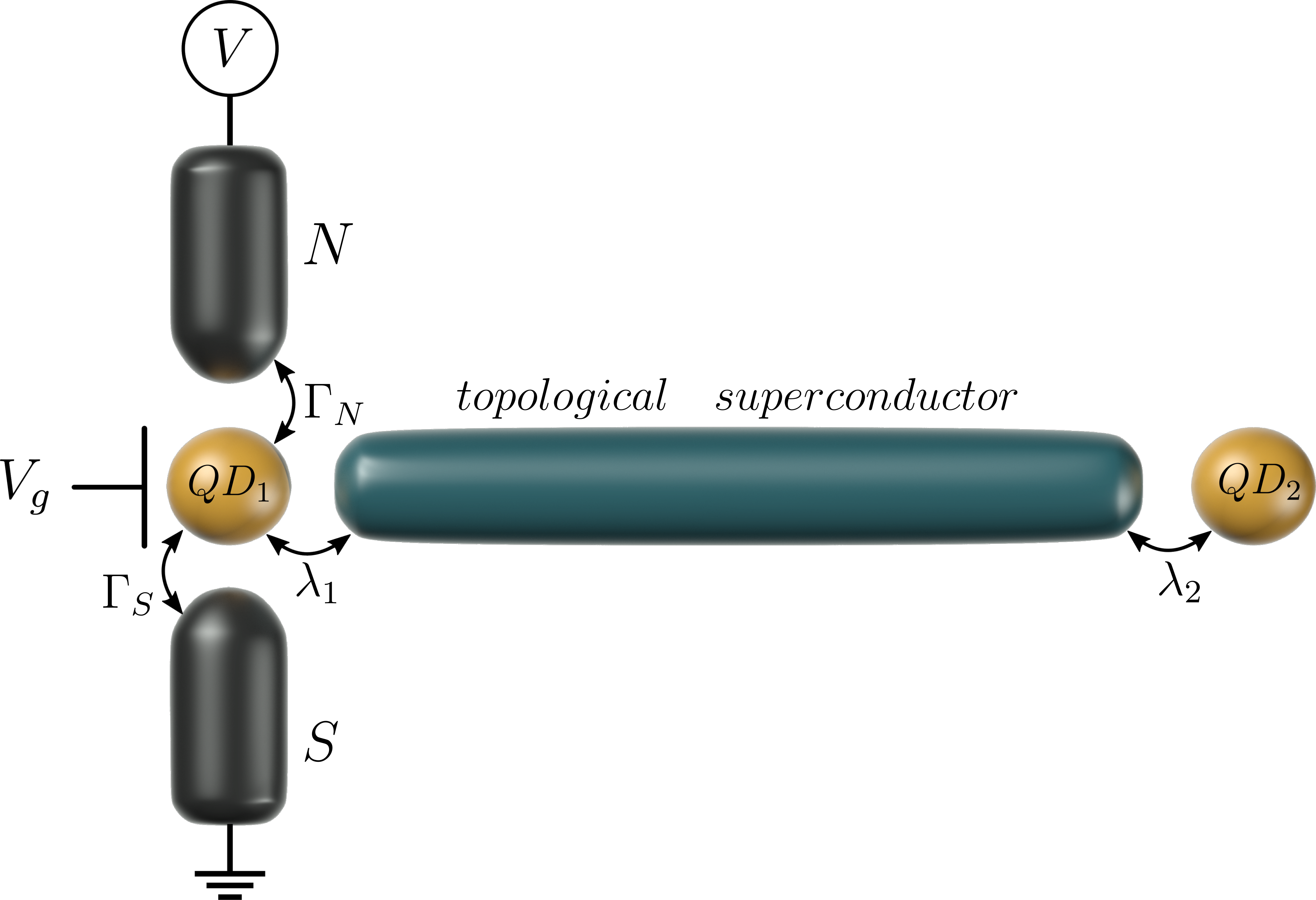}
\caption{Schematic view of two quantum dots attached to the opposite sides  
of the topological superconducting nanowire, hosting Majorana boundary 
modes. The first quantum dot, QD$_{1}$, is placed between the normal lead (N) and conventional 
superconductor (S), so that its emerging quasiparticles can 
be probed by the Andreev spectroscopy. The second quantum dot, QD$_{2}$, is floating.}
\label{scheme}
\end{figure}

Specifically, we investigate the dynamical properties of our setup (displayed 
in Fig.\ \ref{scheme}) right after coupling its constituents. The system 
consists of two quantum dots hybridized with the topological superconducting 
nanowire, hosting the Majorana boundary modes. One of the quantum dots, QD$_{1}$, 
is embedded between the conventional superconductor (S) and normal (N) leads,  
enabling its quasiparticles to be probed by the Andreev (electron-to-hole 
scattering) spectroscopy. The second quantum dot, QD$_{2}$, is flanked on the opposite side 
of the topological superconductor. These spatially distant quantum dots are 
communicated solely through the Majorana zero-energy modes of the topological 
superconductor. In what follows, we inspect nonlocal phenomena, appearing 
in the transient dynamics of measurable observables.

It is important to note that the time-resolved studies of topological superconductors
have so far addressed various aspects, including 
their dynamics imposed by quantum quench across the topological transition 
in the Rashba nanowires \cite{Tuovinen-2021,Tuovinen_2019}, the nonequilibrium effects 
induced upon switching on/off the topological phase in segments of 
the Kitaev chains \cite{Dagotto-2023,Alicea-2018}, gradual leakage 
of the Majorana mode onto single quantum dot \cite{Baranski-2021},
the crossed-Andreev and ballistic charge transfer processes
\cite{Li_2020}, nonequilibrium dynamics of the Majorana-Josephson system 
\cite{Braunecker_2018}, as well as the  waiting times of a topological Andreev interferometer \cite{Dutta_2024}. As regards the nonlocal effects, they have 
been mainly investigated under the static conditions, considering 
finite hybridization of the Majorana modes \cite{Deng-2018}.
Our study is therefore complementary to those ones, focusing on the nonlocal 
dynamical effects detectable in the hybrid structures with topological superconducting nanowire.

For microscopic considerations we assume the constituents of our setup 
to be disconnected till $t=0$. After coupling them, we explore the 
quantum evolution of physical observables (for $t>0$). In 
particular, we determine: the charge occupancy of both quantum dots, 
the local and nonlocal electron pairings and the charge current 
flowing through QD$_{1}$ in the unbiased and biased setup. The differential
conductance of such current could enable empirical detection of 
the gradually emerging trivial and topological bound states of 
QD$_{1}$. We find that nonlocal effects extend solely 
over the transient region and later on (when additional abrupt or 
continuous changes are imposed on the energy levels of 
quantum dots and/or the coupling to external leads) they completely
vanish.

The paper is organized as follows. In Sec.~\ref{sec_model} we formulate
the microscopic model. Section~\ref{dynamics} presents the method relevant to
uncorrelated system and discusses the analytical and numerical results obtained 
in the transient region of the unbiased setup. Next, in Sec.\ \ref{biased_system}, 
we show the numerical results for the time-dependent charge transport induced
by the voltage applied across QD$_{1}$ between the external N/S leads. 
In Sec.~\ref{correlations} we study the competition of the Coulomb repulsion
with the superconducting proximity effect, manifested in the local and nonlocal 
electron pairings, by means of the time-dependent numerical renormalization group method. Section~\ref{summary} concludes the paper and summarizes the main findings. Appendices 
\ref{app} and \ref{app_B} present useful technical details, concerning derivation 
of the time-dependent physical observables.  

\section{Microscopic model}
\label{sec_model}

The hybrid structure displayed in Fig.\ \ref{scheme} can  be modeled 
by the following Hamiltonian
\begin{equation}
\hat{H} =   \sum_{i=1,2} \hat{H}_{QD_{i}}  +
\sum_{\beta=N,S} \big( \hat{H}_{\beta} + 
\hat{H}_{QD_{1}-\beta} \big) +  \hat{H}_{M-DQD} .
\label{model}
\end{equation}
The $i$-th quantum dot (QD$_{i}$) is treated as the Anderson-type 
impurity
\begin{eqnarray}
\hat{H}_{QD_{i}}=\sum_{\sigma} 
\varepsilon_{i} \hat{d}^{\dagger}_{i\sigma} 
\hat{d}_{i \sigma} + U\hat{n}_{i\uparrow} \hat{n}_{i\downarrow} ,
\end{eqnarray}
where the operator $\hat{d}_{i\sigma}$ ($\hat{d}^{\dagger}_{i\sigma}$) annihilates 
(creates) electron with spin $\sigma=\uparrow, \downarrow$ at the energy level 
$\varepsilon_{i}$ and $U$ is the  Coulomb repulsion between opposite spin electrons. 

We assume that QD$_{1}$ is embedded between the superconducting ($\beta=S$) and 
normal ($\beta=N$) leads. The superconductor is described by the BCS-type Hamiltonian
%
$\hat{H}_{S} =\sum_{{\bf k},\sigma}  \xi_{S {\bf k}} \hat{c}_{S {\bf k}\sigma}^{\dagger}
\hat{c}_{S{\bf k}\sigma} \!-\! \sum_{\bf k} \big( \Delta_{SC}   \hat{c}_{S{\bf k} \uparrow} ^{\dagger}
\hat{c}_{S -{\bf k} \downarrow}^{\dagger} + \mbox{H.c.} \big)$,
%
where the pairing gap $\Delta_{SC}$ is isotropic and the energies $\xi_{S{\bf k}}=
\varepsilon_{S{\bf k}}-\mu_{S}$ are expressed with respect to the chemical potential $\mu_{S}$. 
The normal lead is treated as a free fermion gas of itinerant electrons, $\hat{H}_{N} \!=\! \sum_{{\bf k},\sigma} 
\xi_{N {\bf k}} \hat{c}_{N{\bf k} \sigma}^{\dagger}\hat{c}_{N{\bf k} \sigma}$, where 
$\xi_{N{\bf k}}=\varepsilon_{N{\bf k}}-\mu_{N}$. Practically, the latter can be thought of as
a metallic tip of the scanning tunneling microscope (STM). External voltage, $V$, applied 
across the leads detunes the chemical potentials, $\mu_{N}-\mu_{S}=eV$, inducing the charge 
current. Such processes arise from the hybridization terms  
\begin{eqnarray}
\hat{H}_{QD_{1}-\beta} = \sum_{{\bf k},\sigma} \big( V_{\beta {\bf k}} \; \hat{d}_{1\sigma}^{\dagger}
\hat{c}_{\beta {\bf k} \sigma} + V^{*}_{\beta {\bf k}} \; \hat{c}_{\beta {\bf k} \sigma}^{\dagger}
\hat{d}_{1\sigma} \big),
\end{eqnarray}
where $V_{\beta {\bf k}}$ denote the corresponding tunnel matrix elements. We restrict our
considerations to the Andreev (electron-to-hole) scattering regime, which occurs for 
small voltages, $|eV| < \Delta_{SC}$. Under such circumstances one can use the 
wide-band limit approximation, introducing the constant couplings, $\Gamma_{\beta} = \pi\sum_{\textbf{k}}|V_{\beta\textbf{k}}|^{2} \delta(\varepsilon-
\varepsilon_{\beta\textbf{k}})$. 

Focusing our study on transient phenomena in the subgap regime, we shall treat 
the pairing gap $\Delta_{SC}$ as the largest energy scale. Superconducting proximity 
effect can be then modeled by 
$\hat{H}_{S}+\hat{H}_{QD_{1}-S} \approx  {\Gamma_{S} \big( 
\hat{d}_{1\downarrow}^{\dagger}\hat{d}_{1\uparrow}^{\dagger} + 
\hat{d}_{1\uparrow}\hat{d}_{1\downarrow} \big)}$,
where $\Gamma_{S}$ plays the role of the electron pairing 
induced at QD$_{1}$ \cite{Oguri-2004}.

The last term in (\ref{model}) describes the Majorana modes of 
the topological superconducting nanowire \cite{Baranger-2011,Flensberg-2011,Liu-2014}
\begin{eqnarray}
\hat{H}_{M-DQD} &=&  \lambda_{1} \big( \hat{d}^{\dagger}_{1\uparrow} -
\hat{d}_{1\uparrow} \big) \hat{\gamma}_{1} + i\lambda_{2} \hat{\gamma}_{2} 
\big( \hat{d}_{2\uparrow}^{\dagger} 
+  \hat{d}_{2\uparrow} \big) \nonumber \\
&+& i \varepsilon_M \hat{\gamma}_1 \hat{\gamma}_2 ,
\label{M-DQD}
\end{eqnarray}
where $\hat{\gamma}^{\dagger}_i=\hat{\gamma}_{i}$ are self-hermitian Majorana operators.
We assume that only the spin-$\uparrow$ electrons of the quantum dots are coupled 
to these Majorana boundary modes with the coupling strength $\lambda_{i}$. For 
nanowires shorter than the superconducting coherence length  one should take 
into account an overlap $\varepsilon_M$ between the Majorana modes. Here we restrict ourselves
to sufficiently long nanowires, where such overlap is practically negligible, unless stated otherwise.

For convenience, we recast the self-hermitian operators $\hat{\gamma}_i
=\hat{\gamma}_i^{\dagger}$ by the operators $\hat{f}_i^{(\dagger)}$ 
defined through the Bogoliubov transformation 
\begin{eqnarray}
\hat{\gamma}_1 & = & \frac{1}{\sqrt{2}} (\hat{f}^\dagger+\hat{f}), \\
\hat{\gamma}_2 & = & \frac{i}{\sqrt{2}} (\hat{f}^\dagger-\hat{f}) ,
\end{eqnarray}
which obey the anticommutation relations $\big\{ \hat{f},
\hat{f}^{\dagger}\big\}=1$ of the conventional fermion fields.

\section{Dynamics of unbiased setup}
\label{dynamics}

Let us first study the time-dependent physical observables of our unbiased 
hybrid structure, neglecting the correlations, $U=0$. For this purpose we 
adapt the method introduced earlier in Refs. \cite{Baranski-2021,Wrzesniewski-2021,
Taranko-2018}. Specifically, we solve the Heisenberg equations of motion 
$i\hbar \partial_{t} \hat{O}  = \big[ \hat{O},\hat{H} \big]$ for the 
localized $\hat{d}_{i\sigma}^{(\dagger)}(t)$ and itinerant 
$\hat{c}_{\beta {\bf k}\sigma}^{(\dagger)}(t)$ electrons. 

We assume the components of our setup (Fig. \ref{scheme}) to be 
disconnected till $t\leq 0$. This implies that initial ($t=0$) expectation 
values of the mixed operators vanish, i.e.
$\expect{ \hat{d}_{i\sigma}^{(\dagger)}
(0) \hat{d}_{j\neq i \sigma'}(0) } =0$ \cite{Bondyopadhaya-2019}. 
The system is next abruptly formed and we study its evolution for $t > 0$ 
solving the coupled differential equations of the second quantization
operators. It is convenient to introduce the Laplace transforms 
$O(s) = \int_{0}^{\infty} e^{-st}\hat{O}(t)dt$ to account 
for the initial conditions of arbitrary physical observables. This 
approach is reliable for uncorrelated structures, $U=0$, but it can 
be also generalized by incorporating the perturbative treatment of 
interactions. Finally, the time-dependent observables $O(t)$ can be 
determined from the inverse Laplace transforms ${\cal{L}}^{-1}
\left\{O(s)\right\}(t)$.

In what follows, we assume the energy levels of both quantum dots to
be equal, $\varepsilon_{i}=0$, for which we derive analytical expressions 
of the Laplace transforms for $\hat{d}_{i\sigma}^{(\dagger)}(s)$ 
and $\hat{f}^{(\dagger)}(s)$. Their inverse Laplace transforms yield 
analytical expressions for the expectation values of various local 
and nonlocal observables that are of interest in this paper.  
We also note that in the absence of correlations on the quantum dots the operator
$\hat{d}_{2\downarrow}^{(\dagger)}(s)$ decouples from the system dynamics,
therefore, we do not consider it in sections \ref{dynamics}-\ref{biased_system}. 

\subsection{Laplace-transformed equations of motion} 

Heisenberg equations of motion for the localized electrons  
$\hat{d}_{1\sigma}^{(\dagger)}$ mix them (via the hybridization $V_{\beta{\bf k}}$) 
with equations for the itinerant electrons $\hat{c}_{\beta {\bf k}\sigma}^{(\dagger)}$. 
Their detailed derivation has been previously discussed by us in Refs
\cite{Taranko-2018,Wrzesniewski-2021}, considering a single quantum dot conventional N-QD-S nanostructure. In the present case, however, 
we must additionally take into account the operators $\hat{f}^{(\dagger)}(s)$ 
originating from the coupling $\lambda_{1}$ \cite{Baranski-2021} and 
indirectly  also the operators $\hat{d}_{2\uparrow}^{(\dagger)}$ 
because of the coupling $\lambda_{2}$. 

After some algebra, we find the following Laplace transforms 
(valid for $\varepsilon_{i}=0$) 
\begin{eqnarray}
\hat{d}_{1\uparrow}(s)&=& \hat{d}_{1\uparrow}^{\dagger}(0)
\frac{\lambda_{1}^{2}(s+\Gamma_{N})^{2}}{H_{3}(s)} + \hat{d}_{1\uparrow}(0)
\frac{(s+\Gamma_{N})H_{1}(s)}{H_{3}(s)} \label{d1up}
\nonumber \\
&-&i \hat{d}_{1\downarrow}^{\dagger}(0) \frac{\Gamma_{S} H_{1}(s)}{H_{3}(s)}
+i\hat{d}_{1\downarrow}(0) \frac{\Gamma_{S}\lambda_{1}^{2}(s+\Gamma_{N})}{H_{3}(s)}
\label{eq7}
\\
&-&i\big[ \hat{f}(0) + \hat{f}^{\dagger}(0)\big]
\frac{\lambda_{1}(s+\Gamma_{N})}{\sqrt{2}H_{2}(s)}+\sum_{\bf k} V_{{\bf k}}\hat{S}_{\bf k}(s) ,
\nonumber \\
\hat{d}_{2\uparrow}(s)&=& \hat{d}_{2\uparrow}(0)
\frac{s^{2}+\lambda_{2}^{2}}{s(s^{2}+2\lambda_{2}^{2})} 
- \hat{d}_{2\uparrow}^{\dagger}(0)\frac{\lambda_{2}^{2}}{s(s^{2}+2\lambda_{2}^{2})} 
 \label{d2up}
\nonumber \\
&+&i \big[ \hat{f}(0) - \hat{f}^{\dagger}(0)\big]
\frac{\lambda_{2}}{\sqrt{2}(s^{2}+2\lambda_{2}^{2})} ,
\end{eqnarray}
%
\begin{widetext}
\begin{eqnarray}
\hat{f}(s)&=& i \big[ \hat{d}_{1\uparrow}^{\dagger}(0)-\hat{d}_{1\uparrow}(0) \big] 
\frac{\lambda_{1}(s+\Gamma_{N})}{\sqrt{2}H_{2}(s)} - \big[ \hat{d}_{1\downarrow}^{\dagger}(0)
+\hat{d}_{1\downarrow}(0) \big] \frac{\lambda_{1}\Gamma_{S}}{\sqrt{2}H_{2}(s)}
+i \big[ \hat{d}_{2\uparrow}^{\dagger}(0)+\hat{d}_{2\uparrow}(0) \big]
\frac{\lambda_{2}}{\sqrt{2}(s^{2}+2\lambda_{2}^{2})} 
\nonumber \\
&+& \hat{f}(0) A(s) + \hat{f}^{\dagger}(0) B(s)  
+\frac{i\lambda_{1}}{\sqrt{2}s} \sum_{\bf k} V_{{\bf k}}
\big[\hat{S}_{\bf k}^{\dagger}(s)-\hat{S}_{\bf k}(s)\big],
\label{eq9}
\end{eqnarray}
%
\end{widetext}
\noindent
where 
%
\begin{equation}
H_{1}(s) = s^{3}+2\Gamma_{N}s^{2} +(\Gamma_{N}^{2}+\Gamma_{S}^{2}+\lambda_{1}^{2}) s + \lambda_{1}^{2}\Gamma_{N},
\label{H1}
\end{equation}
\begin{equation}
H_{2}(s) = s^{3}+2\Gamma_{N}s^{2} +(\Gamma_{N}^{2}+\Gamma_{S}^{2}+2\lambda_{1}^{2}) s + 2\lambda_{1}^{2}\Gamma_{N},
\label{H2}
\end{equation}
\begin{equation}
H_{3}(s) = (s^{2}+2s\Gamma_{N} + \Gamma_{N}^{2}+\Gamma_{S}^{2}) H_{2}(s),
\label{H3}
\end{equation}
\begin{equation}
A(s) = \frac{s^{2}+\lambda_{2}^{2}}{s(s^{2}+2\lambda_{2}^{2})} - \frac{\lambda_{1}^{2}(s+\Gamma_{N})}{sH_{2}(s)} ,
\end{equation}
\begin{equation}
B(s) = \frac{s^{2}(\lambda_{2}^{2}-\lambda_{1}^{2})+s\Gamma_{N}(2\lambda_{2}^{2}
-\lambda_{1}^{2}) + \lambda_{2}^{2}(\Gamma_{N}^{2}+\Gamma_{S}^{2})}{(s^{2}+2\lambda_{1}^{2})H_{2}(s)} ,
\label{B_s}
\end{equation}
\begin{eqnarray}
\hat{S}_{\bf k}(s) & = & 
\frac{\hat{c}_{{\bf k}\downarrow}^{\dagger}(0)H_{1}(s)\Gamma_{S} }{H_{3}(s)(s-i\xi_{{\bf k}})}
-i\frac{\hat{c}_{{\bf k}\uparrow}(0)H_{1}(s)(s+\Gamma_{N})}{H_{3}(s)(s+i\xi_{{\bf k}})}
\nonumber \\
&+& \frac{\hat{c}_{{\bf k}\downarrow}(0)\lambda_{1}^{2}(s+\Gamma_{N})\Gamma_{S}}{H_{3}(s)(s+i\xi_{{\bf k}})}
+i\frac{\hat{c}_{{\bf k}\uparrow}^{\dagger}(0)\lambda_{1}^{2}(s+\Gamma_{N})^{2}}{H_{3}(s)(s-i\xi_{{\bf k}})},
\label{S}
\end{eqnarray}
where we now use 
$V_{\bf k}\equiv V_{{N}\bf k}$, $\hat{c}_{{\bf k} \sigma}^{\dagger}\equiv \hat{c}_{N{\bf k} \sigma}^{\dagger}$, $\xi_{\bf k}\equiv \xi_{{N}\bf k}$, and $\varepsilon_{\bf k}\equiv \varepsilon_{{N}\bf k}$.
The Laplace transform of $\hat{d}_{1\downarrow}(s)$
can be obtained from the  exact relation
\begin{equation}
\hat{d}_{1\downarrow}(s) = \frac{1}{s+\Gamma_{N}} \left[ i\Gamma_{S} \hat{d}^{\dagger}_{1\uparrow}(s)
+ \hat{d}_{1\downarrow}(0) -i \sum_{\bf k} \frac{V_{{\bf k}}\hat{c}_{{\bf k}\downarrow}(0)}
{s+i \xi_{{\bf k}}} \right] .  \label{d1down}
\end{equation}
We can notice that $\hat{d}_{1\sigma}^{(\dagger)}(s)$ does not depend on the second 
quantum dot operators $\hat{d}_{2\sigma'}^{(\dagger)}(0)$. Similarly, the operator $\hat{d}_{2\uparrow}^{(\dagger)}(s)$
neither depends on $\hat{d}_{1\sigma'}^{(\dagger)}(0)$ 
nor on $\hat{c}_{{\bf k}\sigma'}^{(\dagger)}(0)$.
Such properties are shown here
explicitly for $\varepsilon_{i}=0$, but they are valid for arbitrary energy 
levels as well. On this basis, one can expect that physical observables corresponding
to different quantum dots should be independent of one another. For instance, the charge 
occupancy of QD$_{1}$ should neither depend on the coupling $\lambda_{2}$ nor 
the energy level $\varepsilon_{2}$. 
In other words, the charges accumulated at the quantum dots at a given time 
instant $t>0$ are expected to be uncorrelated \cite{Feng_2021}. In the remaining
part of this section we shall check whether such expectation is really true or not. 

Using the inverse Laplace transforms of  $\hat{d}_{i\sigma}^{(\dagger)}(s)$ we can 
explicitly determine the charge occupancy at each quantum dot $n_{i\sigma}(t)=\langle \hat{d}_{i\sigma}
^{\dagger}(t)\hat{d}_{i\sigma}(t)\rangle$, the induced on-dot $\langle \hat{d}_{i\downarrow}
(t)\hat{d}_{i\uparrow}(t)\rangle$, inter-dot $\langle \hat{d}_{1\downarrow}(t)
\hat{d}_{2\uparrow}(t)\rangle$ electron pairings, etc. Another quantity of our interest 
will be the charge current $j_{N\sigma}(t)$ flowing from the normal lead to QD$_{1}$  
because (in presence of the external voltage $eV=\mu_{N}-\mu_{S}$) its differential 
conductance $G_{\sigma}(V,t)=\partial j_{N\sigma}(t)/\partial V$ could 
empirically probe the dynamically-evolving quasiparticles of QD$_{1}$ 
(see Sec.\ \ref{biased_system}).

For convenience, we assume the superconducting lead to be grounded, $\mu_{S}=0$. 
To simplify notation, we set $\hbar=e=k_{B}=1$ and use the coupling $\Gamma_{S}$ 
as a unit for the energies, unless stated otherwise.
In this convention, time will be expressed in 
units of $\hbar/\Gamma_{S}$ and the currents in units of $e\Gamma_{S}/\hbar$, respectively. In realistic situations $\Gamma_{S}\sim 200$ $\mu$eV, 
therefore typical time-scales would of the order of $\sim 3.3$ ps and the characteristic current unit would be $\sim 48$ nA.

\subsection{Time-dependent occupations} 
\label{charge_of_double_dots}

We start by investigating the time-dependent occupation number $n_{i\sigma}(t)$ 
of the quantum dots and another expectation value $n_{f}(t)=\langle 
\hat{f}^{\dagger}(t)\hat{f}(t)\rangle$, related with the Majorana quasiparticles. 
Below, we present explicit expression for the spin-$\uparrow$ occupancy of 
QD$_{1}$ obtained for $\varepsilon_{i}=0$ (its detailed derivation 
is presented in Appendix \ref{app}). Using the Laplace transform 
(\ref{d1up}), we find 
\begin{widetext}
\begin{eqnarray}
n_{1\uparrow}(t) = M(t)+\sum_{\sigma} n_{1\sigma}(0) M_{\sigma}(t) 
+ \frac{\Gamma_{N}}{\pi} \int_{-\infty}^{\infty} d\varepsilon \Phi_{1}(\varepsilon,t)
-\frac{\Gamma_{N}}{\pi} \int_{-\infty}^{\infty} d\varepsilon f_{N}(\varepsilon) 
\left[  \Phi_{1}(\varepsilon,t) - \Phi_{2}(\varepsilon,t) \right]
\label{n_1up}
\end{eqnarray}
where 
%
\begin{eqnarray}
M_{\uparrow}(t) &=& \left( {\cal{L}}^{-1}\left\{ \frac{(s+\Gamma_{N})H_{1}(s)}{H_{3}(s)} \right\}(t) \right)^{2}
- \lambda_{1}^{4} \left( {\cal{L}}^{-1}\left\{ \frac{(s+\Gamma_{N})^{2}}{H_{3}(s)} \right\}(t) \right)^{2} ,
\label{M_up}
\\
M_{\downarrow}(t) &=& \Gamma_{S}^{2} \lambda_{1}^{4}\left( {\cal{L}}^{-1}\left\{ \frac{(s+\Gamma_{N})}{H_{3}(s)} \right\}(t) \right)^{2}
- \Gamma_{S}^{2} \left( {\cal{L}}^{-1}\left\{ \frac{H_{1}(s)}{H_{3}(s)} \right\}(t) \right)^{2} ,
\\
M(t) &=& \Gamma_{S}^{2}\left( {\cal{L}}^{-1}\left\{ \frac{H_{1}(s)}{H_{3}(s)} \right\}(t) \right)^{2}
+ \lambda_{1}^{4} \left( {\cal{L}}^{-1}\left\{ \frac{(s+\Gamma_{N})^{2}}{H_{3}(s)} \right\}(t) \right)^{2}
+ \frac{\lambda_{1}^{2}}{2} \left( {\cal{L}}^{-1}\left\{ \frac{(s+\Gamma_{N})}{H_{2}(s)} \right\}(t) \right)^{2} ,
\label{M_coefficient}
\\
\Phi_{1}(\varepsilon,t) &=& \Gamma_{S}^{2}\left| {\cal{L}}^{-1}\left\{ \frac{H_{1}(s)}{(s+i\varepsilon)H_{3}(s)} \right\}(t) \right|^{2}
+ \lambda_{1}^{4} \left| {\cal{L}}^{-1}\left\{ \frac{(s+\Gamma_{N})^{2}}{(s+i\varepsilon)H_{3}(s)} \right\}(t) \right|^{2} ,
\\
\Phi_{2}(\varepsilon,t) &=& \left| {\cal{L}}^{-1}\left\{ \frac{(s+\Gamma_{N})H_{1}(s)}{(s+i\varepsilon)H_{3}(s)} \right\}(t) \right|^{2}
+ \lambda_{1}^{4} \Gamma_{S}^{2} \left| {\cal{L}}^{-1}\left\{ \frac{(s+\Gamma_{N})}{(s-i\varepsilon)H_{3}(s)} \right\}(t) \right|^{2}
\end{eqnarray}
%
\end{widetext}
\noindent
and $f_{N}(\varepsilon)=\left[1+\exp{\left(\varepsilon/T\right)}\right]^{-1}$ is the Fermi-Dirac distribution function,with $T$ being the temperature and $k_{B}\equiv 1$. In calculations we assume $T\to 0$.
The opposite spin occupancy $n_{1\downarrow}(t)$ can be obtained in the same 
manner [see Eq.\ (\ref{A3}) in Appendix \ref{app}]. Analytical determination 
of $n_{1\downarrow}(t)$ is here feasible, because all needed inverse Laplace 
transforms ${\cal{L}}^{-1}\left\{F(s)\right\}(t)$ can be represented in 
a fractional form $F(s)=(s-s_1) ... (s-s_n)/(s-s_{n+1})...(s-s_{m})$, 
where $2n<m$.  

In agreement with the expectations, we notice that $n_{1\uparrow}(t)$ is 
independent of the initial fillings of $n_{2\sigma}(0)$ and $n_{f}(0)$. 
The term $M(t)$ [see the last part of Eq.\ (\ref{M_coefficient})], however, 
contributes some influence of the Majorana modes to $n_{1\uparrow}(t)$ 
because $\hat{d}_{1\uparrow}(t)$ expanded in terms of the electron operators 
at $t=0$ contains non-vanishing contribution proportional to $\hat{f}(0)$ 
and $\hat{f}^{\dagger}(0)$
\begin{eqnarray}
-i\big[ \hat{f}(0)+\hat{f}^{\dagger}(0)\big] \frac{\lambda_{1}}{\sqrt{2}}
{\cal{L}}^{-1}\left\{ \frac{s+\Gamma_{N}}{H_{2}(s)} \right\}(t) .
\label{eq23_new}
\end{eqnarray}
For $\Gamma_{N}=0$, this expression yields
\begin{eqnarray}
\frac{\lambda_{1}^{2}}{2} \frac{\sin{\left(t\sqrt{\Gamma_{S}^{2}+2\lambda_{1}^{2}}\right)}}
{\Gamma_{S}^{2}+2\lambda_{1}^{2}} .
\end{eqnarray}
Such term does not depend on  $n_{f}(0)$, but it reveals influence 
of QD$_{1}$ coupling $\lambda_{1}$ to the topological nanowire. 

The parts which depend on the initial occupancy of QD$_{1}$ are  separated 
from another contribution dependent on the itinerant electrons of the normal 
lead, represented by the last terms of Eq.\ (\ref{n_1up}). Notice, however, 
that some information about coupling with the normal lead enters $M$, 
$M_{\uparrow}$ and $M_{\downarrow}$ through the term with $\Gamma_{N}$.
It is interesting that $n_{1\sigma}(t)$ is independent of any parameter
characterizing  QD$_{2}$ (i.e.\ $\lambda_{2}$, $\varepsilon_{2\uparrow}$,
$n_{2\sigma}(0)$). Such dependence could eventually arise only for 
nonvanishing overlap $\varepsilon_{M}\neq 0$ between the Majorana modes.

Let us now analyze in some detail the case $\Gamma_{N}=0$, for which explicit 
expressions can be derived. Assuming the initial empty fillings of both QDs, 
$n_{i\sigma}(0)=0$, from the general expression (\ref{n_1up})  we obtain 
\begin{eqnarray}
n_{1\uparrow}(t)&=& \frac{1}{2}+ \frac{\Gamma_{S}}{2\sqrt{\Gamma_{S}^{2}+2\lambda_{1}^{2}}}
\sin{\left( t\Gamma_{S} \right)} \sin{\left( t \sqrt{\Gamma_{S}^{2}+2\lambda_{1}^{2}} \right)}
\nonumber \\
&-& \frac{1}{2} \cos{\left( t\Gamma_{S} \right)} \cos{\left( t \sqrt{\Gamma_{S}^{2}+2\lambda_{1}^{2}} \right)}
\label{eqn17} \\
n_{1\downarrow}(t)&=& \frac{1}{2}+ \frac{\Gamma_{S}}{2\sqrt{\Gamma_{S}^{2}+2\lambda_{1}^{2}}}
\sin{\left( t\Gamma_{S} \right)} \sin{\left( t \sqrt{\Gamma_{S}^{2}+2\lambda_{1}^{2}} \right)}
\nonumber \\
&-& \frac{\Gamma_{S}^{2}}{2(\Gamma_{S}^{2}+2\lambda_{1}^{2})} \cos{\left( t\Gamma_{S} \right)} 
\cos{\left( t \sqrt{\Gamma_{S}^{2}+2\lambda_{1}^{2}} \right)}
\nonumber \\
&-& \frac{\lambda_{1}^{2}}{\Gamma_{S}^{2}+2\lambda_{1}^{2}} \cos{\left( t\Gamma_{S} \right)} .
\label{eqn18}
\end{eqnarray}
For vanishing $\lambda_{1}$ the occupancy has oscillatory behavior
\begin{equation}
n_{1\sigma}(t)=\sin^{2}{\left( t \Gamma_{S}\right)}
\end{equation}
%
with the time period equal to $\pi/\Gamma_{S}$. For the opposite case 
$\lambda_{1} \gg \Gamma_{S}$, we can rewrite Eqs.~(\ref{eqn17}) and (\ref{eqn18}) in the 
following approximate forms 
\begin{eqnarray}
n_{1\uparrow}(t)& \simeq & \frac{1}{2} \left[ 1 - \cos{\left( t\Gamma_{S} \right) } 
\cos{ \left( \sqrt{2}t \lambda_{1} \right) } \right] ,
\label{eqn19} \\
n_{1\downarrow}(t)& \simeq & \sin^{2}{\left( \frac{t\Gamma_{S}}{2} \right) } .
\label{eqn20}
\end{eqnarray}

The occupancy of QD$_{2}$  behaves quite differently in comparison to $n_{1\sigma}(t)$. 
From Eq.\ (\ref{d2up}) we can notice that the operator $\hat{d}_{2\uparrow}(s)$ is 
not coupled to a continuous spectrum of the normal lead, i.e.\ it does not depend 
on $\Gamma_{N}$. For this reason we get its undamped oscillations. Assuming 
the initial condition $n_{2\uparrow}(0)=0$, we analytically obtain
\begin{eqnarray}
n_{2\uparrow}(t)=\frac{2\lambda_{2}^{2}}{2\lambda_{2}^{2}+\varepsilon_{2}^{2}}
\sin^{2}{\left( \frac{t}{2} \sqrt{2\lambda_{2}^{2}+\varepsilon_{2}^{2}} \right) }
\end{eqnarray}
with the period $2\pi/\sqrt{2\lambda_{2}^{2}+\varepsilon_{2}^{2}}$ and 
the constant amplitude (unaffected by the coupling of QD$_{1}$ to the normal 
lead). Only for a finite overlap between the Majorana modes, $\varepsilon_{M}\neq 0$, 
the relaxation processes could be activated, driving the occupancy of QD$_{2}$ 
towards its steady limit ($t\rightarrow \infty$) value .

Using Eq.\ (\ref{eq9}) for $\hat{f}(s)$, we can determine the time-dependent 
occupancy $n_{f}(t)$. For $\varepsilon_{i}=0$ and $\Gamma_{N}=0$, it is 
given by
\begin{eqnarray}
&&n_{f}(t) = \frac{1}{2}+ \left[ 2n_{f}(0) - 1 \right] 
\cos{\left(\sqrt{2}\lambda_{2}t\right)} \times
\label{eq31} \\&&
\left[ \frac{\lambda_{1}^{2}}{\Gamma_{S}^{2}+2\lambda_{1}^{2}} \cos{\left( t
\sqrt{\Gamma_{S}^{2}+2\lambda_{1}^{2}}\right)} 
+\frac{\Gamma_{S}^{2}}{2\left(\Gamma_{S}^{2}+2\lambda_{1}^{2}\right)}
\right] \nonumber ,
\end{eqnarray}
which turns out to be independent of the initial values of $n_{i\sigma}(0)$. 

\begin{figure}
\includegraphics[width=1\columnwidth]{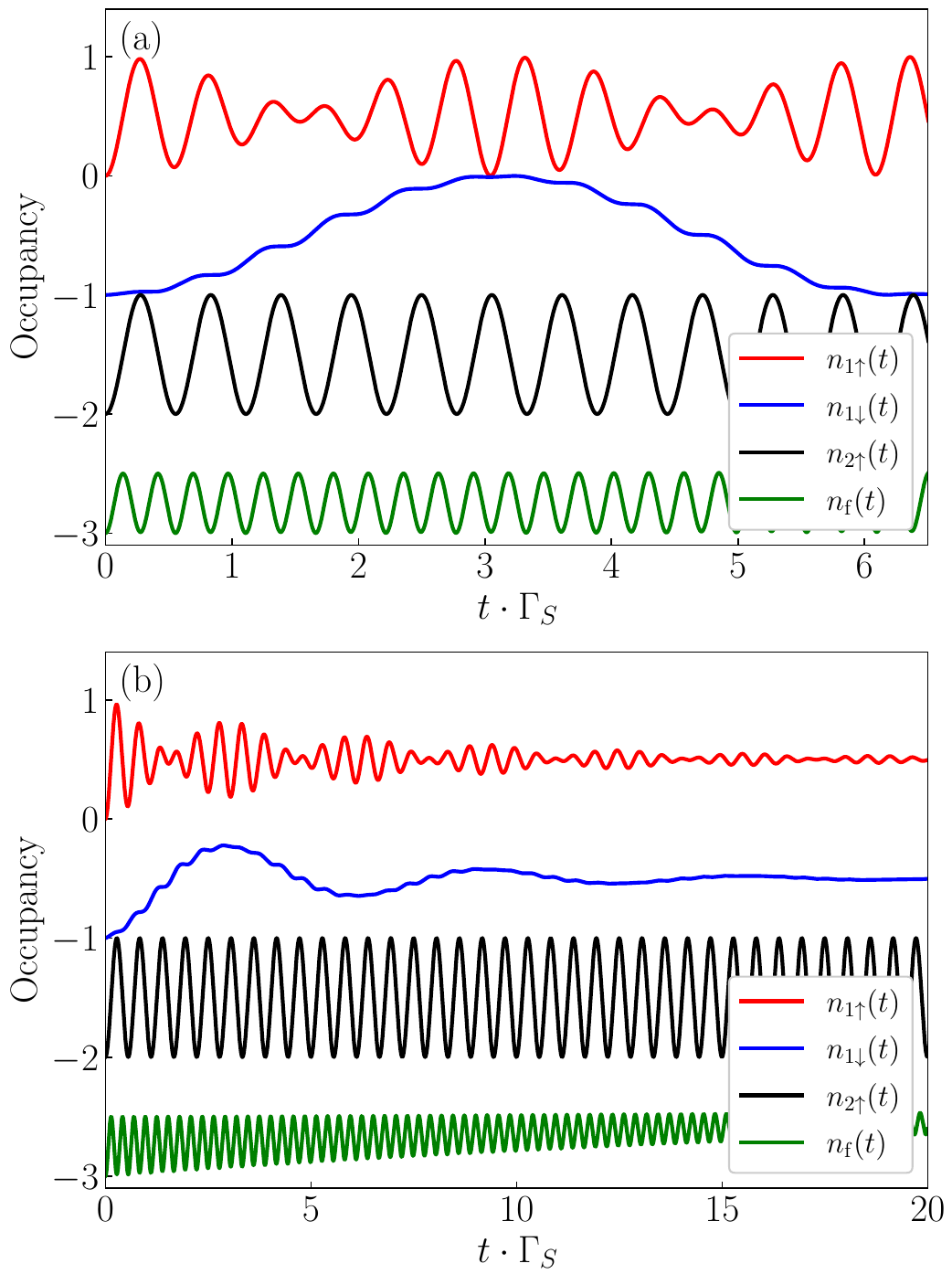}
\caption{Time-dependent occupancies $n_{i\sigma}(t)$ of both quantum dots 
[as indicated] and $n_{f}(t)$ for $\lambda_{1}=\lambda_{2}=8$, 
$\varepsilon_{1}=\varepsilon_{2}=0$, assuming the initial conditions $n_{i\sigma}(0)=n_{f}(0)=0$, 
and $\Gamma_{N}=0$ (upper panel), $\Gamma_{N}=0.2$ (bottom panel).
For clarity the curves are vertically shifted by one (their initial value is zero). All parameters are expressed in units of $\Gamma_S\equiv 1$.}
\label{Fig2}
\end{figure}

Figure \ref{Fig2} presents the time-dependent occupancies 
computed numerically for $\lambda_{1}=\lambda_{2}=8$,
assuming $\Gamma_N=0$ [Fig.~\ref{Fig2}(a)] and 
$\Gamma_N\neq 0$ [Fig.~\ref{Fig2}(b)].
We note that the obtained results nearly coincide 
with the approximate formulas (\ref{eqn19},\ref{eqn20}). The spin-$\downarrow$ 
occupancy of QD$_{1}$ has a similar form as for the $\lambda_{1}=0$ case, but 
the oscillations are twice slower with the period equal to $2\pi/\Gamma_{S}$. Note that generally
the oscillations of $n_{1\sigma}(t)$ are substantially different for each spin component.
In the time-dependent occupancy  $n_{1\uparrow}(t)$ we observe superposition 
of two oscillations: the fast ones, with the period equal to $\sqrt{2} \pi/\lambda_1$, 
and the amplitude oscillations, with the time period equal to  $2\pi/\Gamma_S$ (related 
to the proximitized QD$_{1}$ in absence of the Majorana modes). 
On the other hand, the opposite spin occupancy $n_{1\downarrow}(t)$  oscillates 
with the period equal approximately to  $2\pi/\Gamma_S$,
cf.~Eq.~(\ref{eqn20}).

\begin{figure}[t]
\includegraphics[width=1\columnwidth]{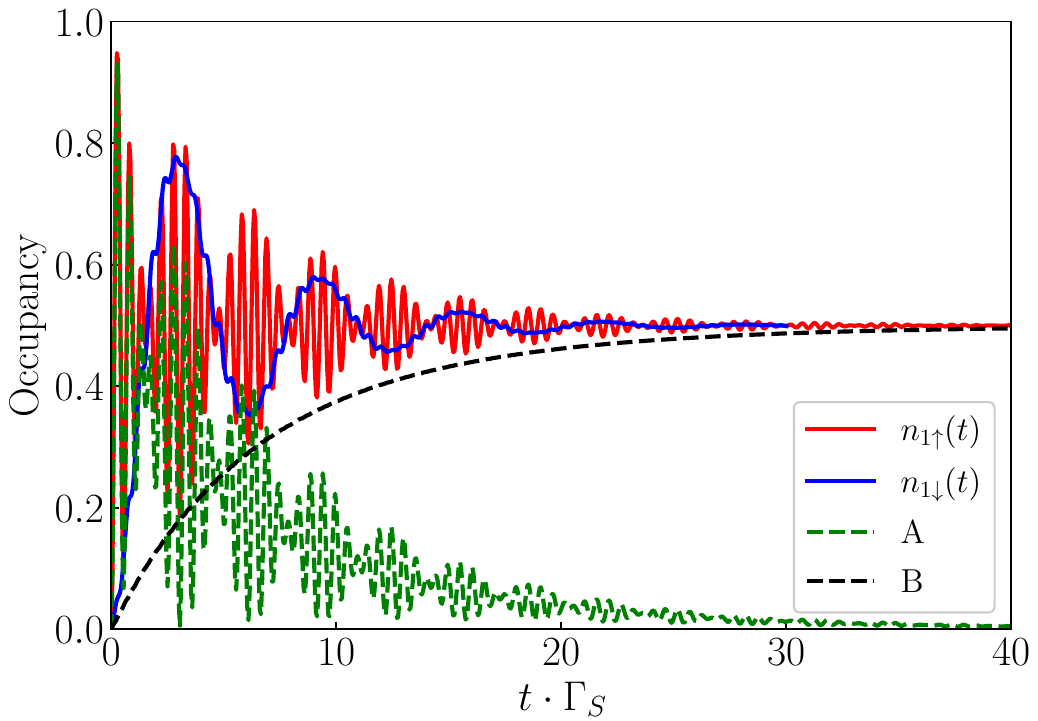}
\caption{The time-dependent occupancies $n_{1\sigma}(t)$ obtained for 
the same set of model parameters as in Fig.\ \ref{Fig2}
with $\Gamma_N=0.2$.
To clarify the behavior of $n_{1\uparrow}(t)$ we show the 
contributions from the term $M(t)$ [curve $A$] and the last term 
of Eq.\ (\ref{eq7}) [curve B].
}
\label{Fig3}
\end{figure}

Let us now analyze the occupation dynamics in the case of finite coupling $\Gamma_N$. The corresponding results are shown in Fig.~\ref{Fig2}(b), while in Fig.~\ref{Fig3} we present detailed behavior of $n_{1\sigma}(t)$ together with contributions stemming from relevant terms of the analytical formulas.
First of all, one can note that now a damping of the oscillatory behavior occurs, see Fig.~\ref{Fig2}(b).
This is even more revealed in different contributions to $n_{1\sigma}(t)$. In particular, the dashed-green line represents the contribution from the term $M(t)$, cf. Eq.~(\ref{M_coefficient}).
One can clearly observe its damped oscillations caused by
the coupling $\Gamma_{N}$.
On the other hand, the black-dashed line represents the contribution due to the coupling of QD$_{1}$ to a continuous spectrum of the normal lead, which basically describes the envelope of the oscillations. The total sum of these contributions
gives the resulting time-dependent occupancy $n_{1\uparrow}(t)$ displayed by the red curve in Fig.~\ref{Fig3}. The stationary limit value $n_{1\uparrow}(t\rightarrow\infty)=0.5$ 
is approached after a sequence of quantum oscillations with the exponentially suppressed amplitude. Performing similar calculations for $n_{1\downarrow}(t)$
we get the result presented by the thick solid line.
Now, the fast oscillations are very much suppressed
and one only observes slow oscillations decaying towards the steady-state value of $1/2$, see the blue curve in Fig.~\ref{Fig3}.

\subsection{Time dependence of the on-dot pairing}
\label{Sec-on-dot}

The time-dependent electron populations of the individual quantum dots are further 
interrelated with development of the local and nonlocal pairings. Let us study 
this issue, first considering the singlet pairing 
\begin{eqnarray}
C_{11}(t) \equiv \expect{ \hat{d}_{1\downarrow}(t) \hat{d}_{1\uparrow}(t) }
\label{intra-dot}
\end{eqnarray}
induced on QD$_{1}$ via its proximity to the (trivial) bulk superconductor.
We can analytically determine $C_{11}(t)$ using the inverse Laplace transforms 
of the operators $\hat{d}_{1\sigma}(s)$ and following the procedure, which is analogous to the
calculations of the charge occupancy $n_{1\sigma}(t)$. For the initially empty 
quantum dots and $\varepsilon_{i}=0$, the on-dot pairing (\ref{intra-dot}) 
is given by 
\begin{equation}
C_{11}(t) = D(t) + i\Gamma_{N}\Gamma_{S}  \int_{-\infty}^{\infty} 
\frac{d\varepsilon}{\pi}
\left[ f_{N}(\varepsilon) D_{1}(\varepsilon,t) + D_{2}(\varepsilon,t) \right]
\label{eq22}
\end{equation}
with functions $D(t)$ and $D_{1,2}(\varepsilon,t)$  presented explicitly 
by equations (\ref{A11}-\ref{A13}) in Appendix \ref{app}.

We notice the absence of any term dependent on $\lambda_{2}$. It means that electron
pairing induced at QD$_{1}$ is completely unaffected by the second QD$_{2}$. 
The pairing function (\ref{eq22}) depends on the initial filling of QD$_{1}$, 
whereas it has no dependence on $n_{f}(0)$, despite the appearance of operators 
$\hat{f}(0)$ and $\hat{f}^{\dagger}(0)$ in the Laplace transform of $\hat{d}_{1\uparrow}(s)$. 
Such terms yield the following result [see the last part of Eq.~(\ref{A11})]  
\begin{equation}
\Big\langle \big[ \hat{f}(0) + \hat{f}^{\dagger}(0) \big]^{2} \Big\rangle 
 \frac{i\Gamma_{S}\lambda_{1}^{2}}{2} 
{\cal{L}}^{-1} \left\{ \frac{1}{H_{2}(s)} \right\}(t)
{\cal{L}}^{-1} \left\{ \frac{s+\Gamma_{N}}{H_{2}(s)} \right\}(t) ,
\end{equation}
which is indeed independent of $n_{f}(0)$.
         
For $\Gamma_{N}=0$  and initially empty QD$_{1}$, the pairing correlation function 
$C_{11}(t)$  is purely imaginary. It can be written in the following simple form       
\begin{eqnarray}
C_{11}(t) &=& -\frac{i}{2} \left[ \frac{\lambda_{1}^{2}}{\Gamma_{S}^{2}+2\lambda_{1}^{2}}
\sin{\left( t\Gamma_{S}\right)} \right. 
\label{eqn23} \\
&+& \frac{\Gamma_{S}^{2}+\lambda_{1}^{2}}{\Gamma_{S}^{2}+2\lambda_{1}^{2}}
\sin{\left( t\Gamma_{S}\right) \cos{\left( t\sqrt{\Gamma_{S}^{2}+2\lambda_{1}^{2}} \right)}} 
\nonumber \\
&+& \left. \frac{\Gamma_{S}}{\sqrt{\Gamma_{S}^{2}+2\lambda_{1}^{2}}}
\cos{\left( t\Gamma_{S}\right) \sin{\left( t\sqrt{\Gamma_{S}^{2}+2\lambda_{1}^{2}} \right)}} 
\right] , 
\nonumber
\end{eqnarray}
representing a combination of the oscillations with frequencies $\Gamma_{S}$ and 
$\sqrt{\Gamma_{S}^{2}+2\lambda_{1}^{2}}$,  respectively.
For $\lambda_{1}\gg \Gamma_{S}$, the oscillations are characterized
by a small period $\sim 2\pi/\sqrt{\Gamma_{S}^{2}+2\lambda_{1}^{2}}$
and the amplitude modulated
by another periodicity of $2\pi / \Gamma_S$.
In this case ${C_{11}(t)\approx -\frac{i}{2}
\sin{\left( t \Gamma_{S}\right)} \cos^{2}{\left( \frac{t}{2} \sqrt{\Gamma_{S}^{2}+2\lambda_{1}^{2}} \right)}}$.

For small values of the coupling $\lambda_{1}$,
one can neglect the terms proportional  
(and of higher order) to $\left( \frac{\lambda_{1}}{\Gamma_{S}}\right)^{2}$ and get 
$C_{11}(t)\approx -\frac{i}{2} \sin{\left( 2 t \Gamma_{S}\right)}$.
This result is identical to the case when QD$_{1}$
is coupled  only to the superconducting lead, 
cf. Refs.~\cite{Taranko-2018,Wrzesniewski-2021}.
For $\Gamma_{N}\neq 0$, the on-dot 
pairing becomes a complex function.
Its real part originates from the last term 
in Eq.~(\ref{d1up}) stemming from continuous spectrum of the normal lead. 
This part depends on the Fermi level of the normal lead,
see Eq.~(\ref{A3}).

\begin{figure}
\includegraphics[width=1\columnwidth]{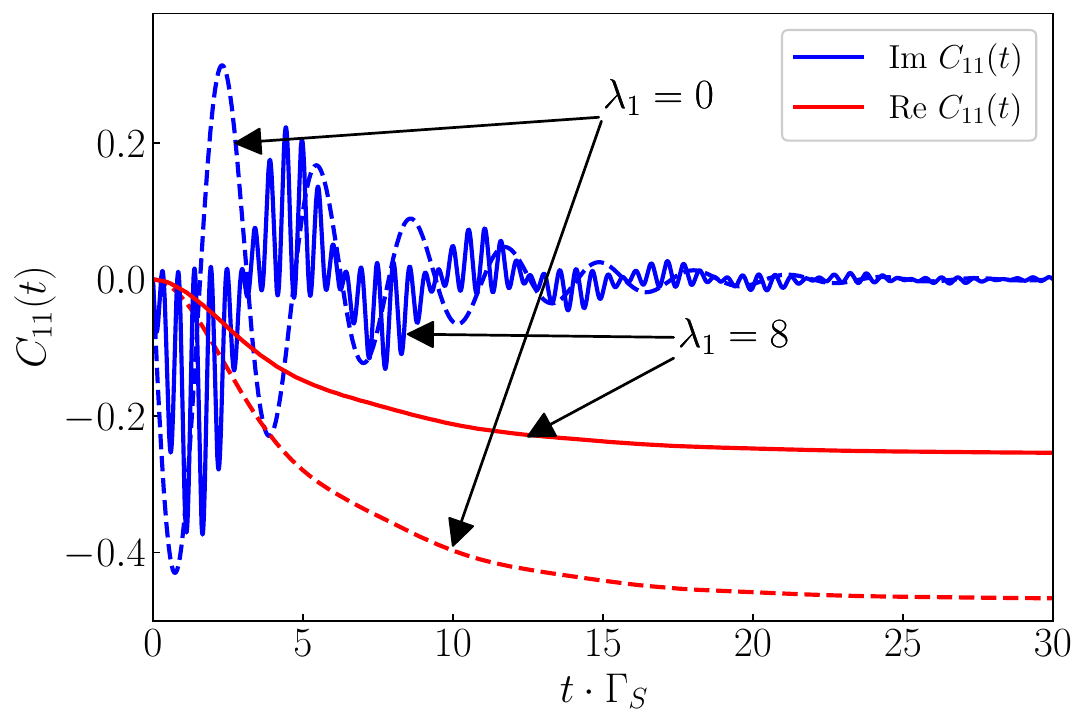}
\caption{The on-dot pairing $C_{11}(t)$ induced at QD$_{1}$ obtained for 
the same set of model parameters as in the bottom panel of Fig.~\ref{Fig2} 
with $\lambda_1=8$ (solid lines). For comparison,
we also show $C_{11}(t)$ in the absence of topological nanowire, i.e. for $\lambda_{1}=0$ (dashed curves).}
\label{Fig4}
\end{figure}

Figure~\ref{Fig4} displays the time-dependent pairing
$C_{11}(t)$ obtained for the unbiased setup ($\mu=0$)
and for finite $\Gamma_N$.
We observe an oscillating structure of the 
imaginary part, similar to the behavior of $n_{1\uparrow}(t)$, cf. Figs.~\ref{Fig2} and \ref{Fig3}.
This is related to the transient charge current, flowing through QD$_{1}$.
These oscillations are damped, because of a continuum states responsible
for the relaxation processes at QD$_{1}$. On the other hand, the 
real part evolves monotonously to its asymptotic (negative) value. For 
comparison, we also plot the complex function $C_{11}(t)$ for the case 
of $\lambda_{1}=0$, i.e.\ when QD$_{1}$ is embedded only between the normal 
and superconducting leads \cite{Taranko-2018}. The structure of
$\mbox{\rm Im} C_{11}(t)$ manifests the high-frequency oscillations
$\sim \sqrt{\Gamma_{S}^{2}+2\lambda_{1}^{2}}$ coexisting with the beats 
of frequency $\sim \Gamma_{S}$. Note that for vanishing $\lambda_{1}$ 
the imaginary part of $C_{11}(t)$ exhibits only one component oscillations 
with the frequency equal to $\Gamma_{S}$.

\begin{figure}
\includegraphics[width=0.95\columnwidth]{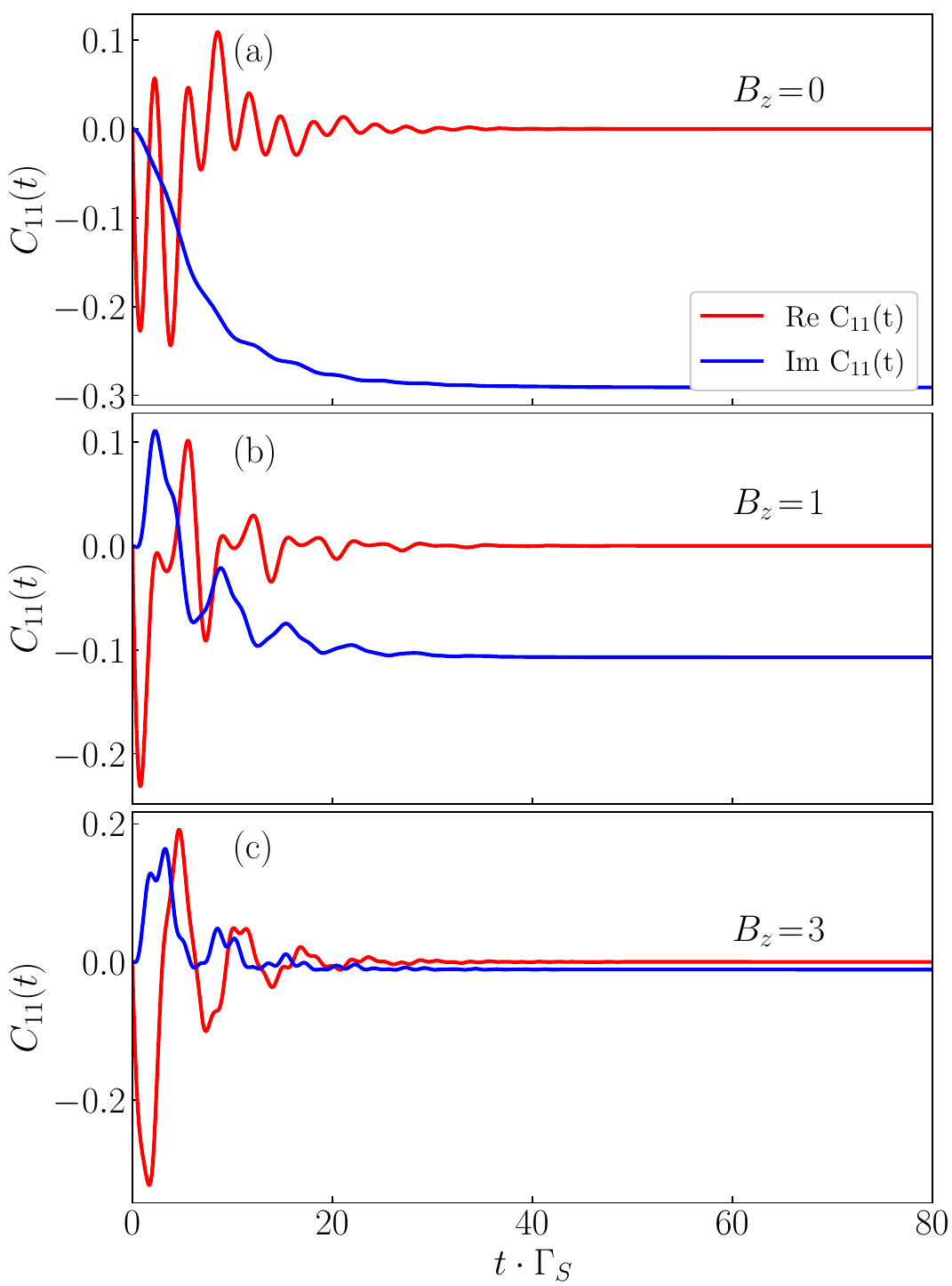}
\caption{
{Transient dynamics of the on-dot pairing $C_{11}(t)$ obtained 
for different values of the magnetic field $B_{z}=0$, $1$, and $3$, using 
the same set of model parameters as in Fig.~\ref{Fig2}.}}
\label{Fig_new_1}
\end{figure}

For hybrid structures, where topological superconductivity is induced by 
contacting the semiconducting nanowires with bulk superconductors, one has
to take into consideration the influence of external magnetic field responsible for the 
Zeeman splitting of the quantum dot energy levels, $B_{z}=\varepsilon_{i\downarrow}
-\varepsilon_{i\uparrow}$. In analogy to a detrimental role of magnetic impurities in bulk superconductors, this Zeeman splitting turns out to suppress the electron 
pairing at QD$_{1}$ (see Fig. \ref{Fig_new_1}).
One can notice that the magnitude of the stationary 
on-dot pairing $|C_{11}(t\rightarrow\infty)|$ gets reduced 
upon increasing the magnetic field. Furthermore, the transient region narrows and the oscillations of the complex function $C_{11}(t)$ acquire rather
complicated profiles, due to transitions between the
Majorana zero-energy mode and the spin-polarized Andreev states.

\subsection{Time-dependent inter-dot pairing}

We now consider the nonlocal electron pairing induced between the spatially distant 
quantum dots. In analogy to the previous subsection, we focus on the singlet pairing  
\begin{equation}     
C_{12}(t) \equiv \expect{ \hat{d}_{1\downarrow}(t) \hat{d}_{2\uparrow}(t) }.\label{interdot_pair}
\end{equation}
In practice this sort of electron pairing could be detected via the crossed 
Andreev reflections in hybrid structures with an additional electrode 
contacted to QD$_{2}$. Let us remark, that $C_{12}(t)$ originates from 
the local pairing of QD$_{1}$ electrons, which is further extended onto 
QD$_{2}$ by the Majorana quasiparticles. Technically this is 
provided by the operators $\hat{f}^{(\dagger)}(0)$ appearing in the Laplace 
transforms of $\hat{d}_{1\downarrow}(s)$ and $\hat{d}_{2\uparrow}(s)$ 
[see Eqs.\ (\ref{d2up}) and (\ref{d1down})]. For $\varepsilon_{i}=0$ we find
\begin{eqnarray}
C_{12}(t) &=& i\lambda_{1}\lambda_{2} \Gamma_{S} \left[ \frac{1}{2} - n_{f}(0) \right] 
\label{eq24} \\ &\times &  
{\cal{L}}^{-1}\left\{ \frac{1}{H_{2}(s)} \right\}(t)
{\cal{L}}^{-1}\left\{ \frac{1}{s^{2}+2\lambda_{2}^{2}} \right\}(t) .
\nonumber
\end{eqnarray}
In the limit of $\Gamma_{N}=0$ the pairing $C_{12}(t)$ simplifies to
\begin{eqnarray}
C_{12}(t) &=& \frac{i\lambda_{1}\Gamma_{S}}{\sqrt{2}(\Gamma_{S}^{2}+2\lambda_{1}^{2})} 
\left[ 1 - 2n_{f}(0) \right] \label{eq25} \\ &\times &  
\sin{\left( \sqrt{2}t\lambda_{2} \right)} \sin^{2}{\left( \frac{t}{2} \sqrt{\Gamma_{S}^{2}
+2\lambda_{1}^{2}} \right)} .
\nonumber
\end{eqnarray}
On the other hand, for finite $\Gamma_{N}$, we obtain
\begin{eqnarray}
C_{12}(t) &=& \frac{i\lambda_{1}\Gamma_{S}}{2\sqrt{2}} 
\left[ 1 - 2n_{f}(0) \right]  \sin{\left( \sqrt{2}t\lambda_{2} \right)} 
\label{eq26}
\\ & \times & 
\frac{ e^{-at}+ e^{-bt}\left[ \frac{a-b}{c} \sin{\left( ct \right)}
-\cos{\left( ct \right)}\right] }{(b-a)^{2}+c^{2}} ,
\nonumber
\end{eqnarray}
where the parameters $a,b,c$ correspond to three roots $s_{i}$ ($i=1,2,3$)
of the cubic equation $H_{2}(s_{i})=0$. One of these roots, say $s_{1}$, is real
and it defines the positive valued parameter $a>0$ via $s_{1}=-a$. The other parameters, 
$b$ and $c$, are related with the conjugated roots $s_{2}=s_{3}^{\star}$. They are 
expressed as $s_{2/3}=-b\pm ic$, where $b>0$. 

Equation (\ref{eq26}) implies that the inter-dot pairing (\ref{interdot_pair}) 
does not depend on the initial occupation $n_{i\sigma}(0)$ of the quantum dots.  
Furthermore, $C_{12}(t)$ has no explicit dependence on the voltage $\mu$ applied across 
QD$_{1}$ between the external leads because the operators $\hat{c}_{{\bf k}\sigma}(0)$ 
do not appear simultaneously in $\hat{d}_{1\downarrow}(t)$ and $\hat{d}_{2\uparrow}(t)$. 
Influence of the normal lead, however, enters indirectly through the roots $s_{1-3}$ 
(where $a$, $b$ and $c$ depend on $\Gamma_{N}$). We observe that for $\Gamma_{N}=0$ 
the amplitude of $C_{12}(t)$ oscillations diminishes upon increasing $\lambda_{1}$, 
whereas for $\Gamma_{N}\neq 0$, we observe a similar behavior, though with damped oscillations. 

In general, the nonlocal pairing (\ref{interdot_pair}) is sensitive to 
the position of the energy levels $\varepsilon_{i}$ of the quantum dots. It is purely 
imaginary only for $\varepsilon_{2}=0$. Otherwise, for 
$\varepsilon_{2} \neq 0$ and arbitrary QD$_{1}$ energy level, the function 
$C_{12}(t)$ becomes complex with non-vanishing real and imaginary parts. 


\begin{figure}[t]
\includegraphics[width=1\columnwidth]{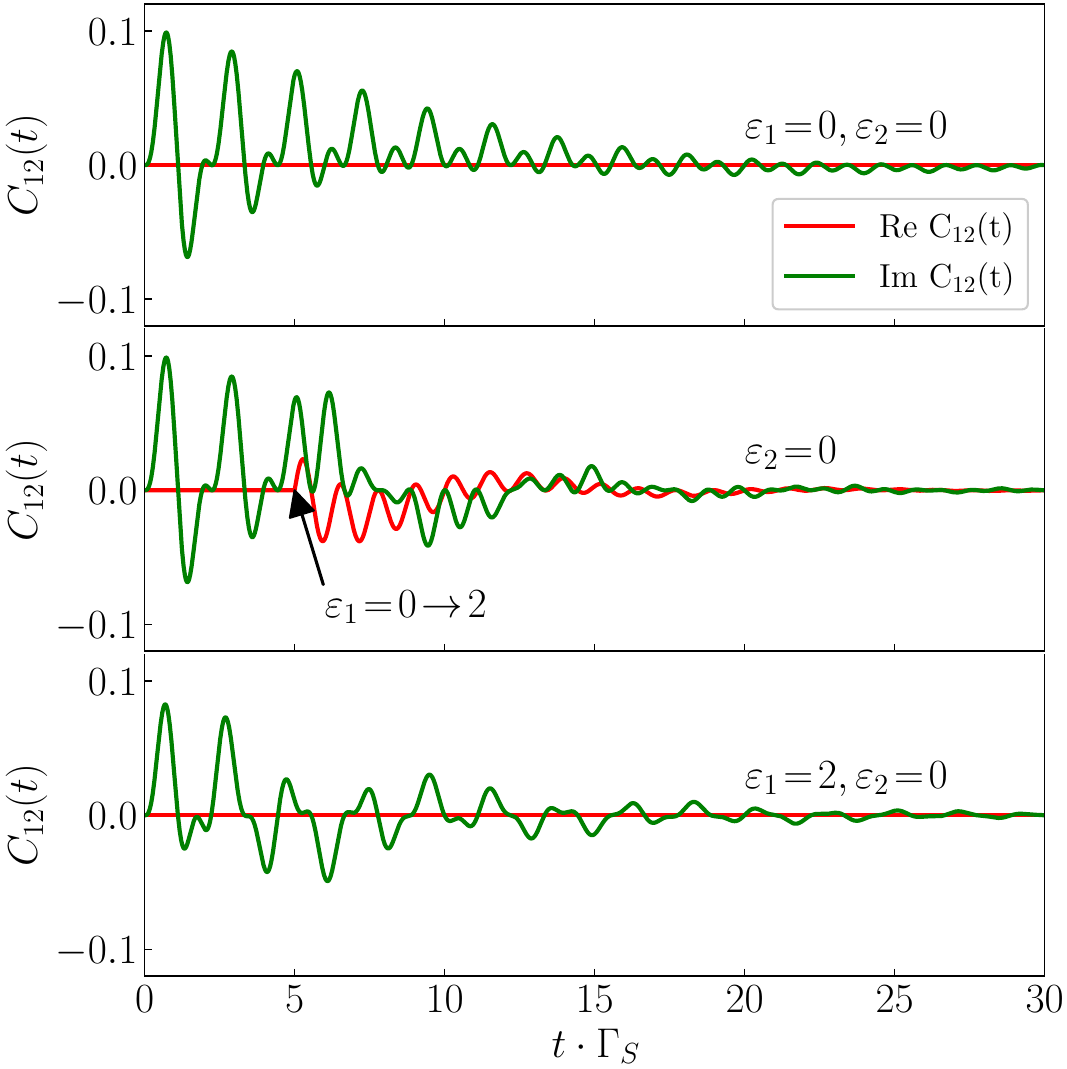}
\caption{The complex inter-dot pairing function $C_{12}(t)$ obtained for 
$\lambda_{1}=\lambda_{2}=2$, $\Gamma_{N}=0.2$ and several values of quantum dot energy levels, as indicated. The red curves refer to the real and the green curves to the imaginary parts of $C_{12}(t)$, respectively. The middle plot presents
the evolution of $C_{12}(t)$ when imposing a sudden change of the energy level 
from $\varepsilon_{1}=0$ to $\varepsilon_{1}=2$ at $t=5/\Gamma_S$.}
\label{Figx}
\end{figure}


The time-dependent inter-dot pairing for selected values of the 
quantum dot energy levels is shown in Fig.~\ref{Figx}.
First of all, we notice that the inter-dot pairing survives only in the transient region. The time-dependent  profile of this complex pairing 
function $C_{12}(t)$ is sensitive to specific values of the quantum dot energy levels. 
Additional effects can arise from the quantum quenches. The middle plot 
in Fig.\ \ref{Figx} displays the evolution of $C_{12}(t)$ after a sudden 
change of $\varepsilon_{1}$ imposed at $t=5/\Gamma_S$. This quench substantially
affects the real and imaginary parts of the nonlocal pairing $C_{12}(t)$. 
It is evident that for $t \geq 5/\Gamma_S$ the time-dependent inter-dot pairing presented in the middle panel of Fig.\ \ref{Figx} differs from the bottom panel,
even though the energy levels are identical.
We have checked, however, that signatures of the nonlocal pairing are 
completely absent in all quantum quenches imposed outside the transient regime.
This observation emphasizes the subtle importance of transient phenomena, 
which enable mutual correlations between spatially distant quantum
dots. Such dynamical non-local effects would be observable in the crossed 
Andreev transmission, and could possibly arise via the single particle 
teleportation as well.

Let us also inspect the influence of the Zeeman field $B_z$ on the nonlocal 
pairing $C_{12}(t)$ between opposite spin electrons.
In analogy to $C_{11}(t)$, we notice that the magnetic field quickly suppresses this inter-dot pairing.
Such effects are visible only in the transient region (because the stationary limit value of $C_{12}(t\rightarrow \infty)$ vanishes for all cases).
This suppression arises from a competition of the magnetic field with the singlet pairing, no matter whether it is local or non-local.

\begin{figure}
\includegraphics[width=0.95\columnwidth]{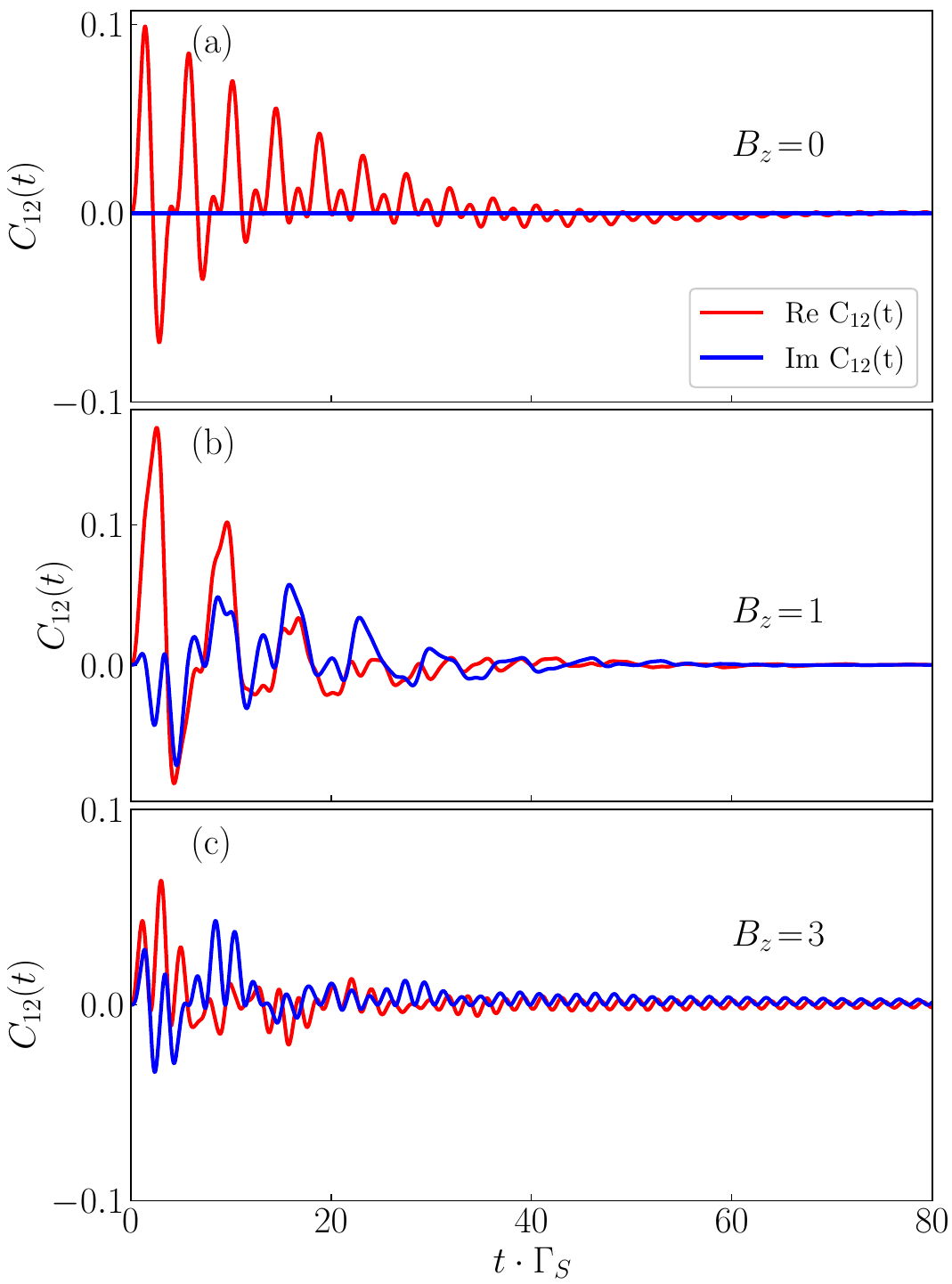}
\caption{
Transient dynamics of the nonlocal pairing $C_{12}(t)$ 
in the singlet channel obtained for the same set of model parameters 
as in Fig.~\ref{Fig_new_1}.}
\label{Fig_new_2}
\end{figure}

\subsection{Dynamics of the triplet pairing}
\label{inter_site_triplet}

Finally let us consider the mixed pairing  
between QD$_{2}$ and the topological superconducting nanowire
\begin{equation}     
C_{2f}(t) = \expect{ \hat{d}_{2\uparrow}(t) \hat{f}(t) } .
\label{mixed_pair}
\end{equation}
This triplet pairing is associated with leakage of the Majorana mode 
onto the side-attached  QD$_{2}$, and similar process occurs on QD$_{1}$ 
regardless of $n_{1\sigma}(0)$. Let us recall, that from both operators 
$\hat{d}_{2\uparrow}(s)$ and $\hat{f}(s)$ only the latter one depends on 
$\hat{c}_{{\bf k}\sigma}^{(\dagger)}(0)$. For this reason the mixed pairing 
(\ref{mixed_pair}) does not depend on the normal lead electrons and thereby 
on the bias voltage $V$.

In what follows, we focus on the initially empty  dot
$n_{2\uparrow}(0)=0$ and $n_{f}(0)=0$, assuming the isotropic
couplings $\lambda_{1}=\lambda_{2}\equiv \lambda$. Under such conditions 
we obtain (for $\varepsilon_{i}=0$) 
\begin{eqnarray}
C_{2f}(t) &=& \frac{i}{2} \sin{\left( \sqrt{2}\lambda t\right)}
\left[ \cos^{2}{\left( \frac{\lambda t}{\sqrt{2}}\right)} \right. \nonumber \\
&+& \left.
\lambda^{2} {\cal{L}}^{-1}\left\{ \frac{s\Gamma_{N}+\Gamma_{S}^{2}+\Gamma_{N}^{2}}{(s^{2}+2\lambda^{2})H_{2}(s)}
\right\}(t) \right] .
\label{eq28}
\end{eqnarray}
For $\Gamma_{N}=0$, formula (\ref{eq28}) simplifies to
\begin{equation}
C_{2f}(t) = \frac{i  \sin{\left( \sqrt{2}\lambda t\right)}}
{2(\Gamma_{S}^{2}+2\lambda^{2})}
\left[ \Gamma_{S}^{2}+2\lambda^{2} \cos^{2}{\left( \frac{\sqrt{\Gamma_{S}^{2}+2\lambda^{2}}}{2} t \right) }
\right] ,
\label{eq29}
\end{equation}
revealing an oscillatory behavior. This analytical result (\ref{eq29}) is valid when the quantum dots interconnected 
through the Majorana quasiparticles are not coupled to a continuous spectrum of the normal lead that
could activate the relaxation processes. It appears, however, that even in
the case of $\Gamma_{N}\neq 0$, the mixed pairing function (\ref{eq28}) oscillates against time 
with non-vanishing amplitude. To observe it, we evaluated the pairing function (\ref{eq28})
in the large time limit 
\begin{eqnarray}
C_{2f}(t) &=& \frac{i}{2}  \sin{\left( \sqrt{2}\lambda t\right)}
\left[  \cos^{2}{\left( \frac{\lambda t}{\sqrt{2}}\right) } 
\right. \nonumber \\ 
&+& 2 \lambda^{2} \Gamma_N \left. \mbox{\rm Re} \left( \frac{e^{i\sqrt{2}\lambda t}s_{16}}
{s_{12}s_{13}s_{14}s_{15}}\right) \right] .
\label{mixed_asymptotic}
\end{eqnarray}
Here $s_{nm}=s_{n}-s_{m}$, and $s_{1}=i\sqrt{2}\lambda$, $s_{2}=-s_{1}$, 
$s_{3}=-a$, $s_{4/5}=-b\pm ic$, $s_{6}=-(\Gamma_{S}^2+\Gamma_{N}^2)/\Gamma_N$, where $s_{3}$, 
$s_{4}$, $s_{5}$ are the roots of the cubic equation $H_{2}(s_{i})=0$ and 
$a$, $b$ are real positive parameters. In the large $t$ limit this pairing 
function is purely imaginary and it has oscillating character. The formula 
presented in Eq.\ (\ref{mixed_asymptotic}) is valid for $\varepsilon_{1}
=\varepsilon_{2}=0$.  Otherwise, the operators $\hat{d}_{2\uparrow}(s)$ 
and $\hat{f}(s)$ depend on the energy levels $\varepsilon_{i}$ 
affecting the pairing function $C_{2f}(t)$. In particular, the influence
of QD$_{1}$ on the triplet pairing $C_{2f}(t)$ is incorporated via 
the operator $\hat{f}(s)$.

\section{Tunneling conductance}
\label{biased_system}

In this section we study the differential conductance $G_{\sigma}(V,t)
=\frac{\partial}{\partial V} j_{N\sigma}(t)$ of the time-dependent current 
$j_{N\sigma}(t)$, flowing between QD$_{1}$ and the normal lead. For convenience 
we express $G_{\sigma}(V,t)$ in units of $2e^{2}/h$ and investigate its 
dependence on the applied bias voltage $V$.
We recall that $eV =\mu_{N}-\mu_{S}$
and the topological superconductor is assumed to be grounded
$\mu_S=0$, such that $eV =\mu_{N}\equiv \mu$.
The tunneling current can be determined 
from the evolution of the total number of electrons in the normal electrode. 
For our setup it can be written as \cite{Taranko-2018}
\begin{equation}
j_{N\sigma}(V,t)=2 \mbox{\rm Im} \sum_{\bf k} V_{\bf k} e^{-i \varepsilon_{\bf k}t}
\expect{ \hat{d}_{1\sigma}^{\dagger}(t) \hat{c}_{{\bf k}\sigma}(0)}
-2\Gamma_{N} n_{1\sigma}(t).
\label{eq31bis}
\end{equation}
Substituting the inverse Laplace transform of $\hat{d}_{1\uparrow}(s)$ [see 
Eq.~(\ref{d1up})] we obtain derivative  of the first term appearing in 
Eq.~(\ref{eq31bis}) with respect to $\mu$ in the following way
%
\begin{widetext}
\begin{eqnarray}
&&\frac{\partial}{\partial \mu} 2\mbox{\rm Im} \left( i \sum_{{\bf k},{\bf k}_{1}}
V_{\bf k} V_{{\bf k}_{1}} e^{-i \varepsilon_{\bf k}t} {\cal{L}}^{-1}
\left\{ \frac{(s+\Gamma_{N})H_{1}(s)}{(s-\varepsilon_{{\bf k}_{1}})H_{3}(s)} \right\}(t)
\expect{ \hat{c}_{{\bf k}_{1}\uparrow}^{\dagger}(t) \hat{c}_{{\bf k}\uparrow}(0)}
\right) \nonumber \\
&=& \frac{\partial}{\partial \mu} \frac{2\Gamma_{N}}{\pi} \mbox{\rm Re} \int_{-\infty}^{\infty} d\varepsilon 
f_{N}(\varepsilon) e^{-i\varepsilon t} {\cal{L}}^{-1} \left\{ \frac{(s+\Gamma_{N})H_{1}(s)}
{(s-\varepsilon)H_{3}(s)} \right\}(t) = \frac{2\Gamma_{N}}{\pi} \mbox{\rm Re} 
\left( e^{-i\mu t} {\cal{L}}^{-1} \left\{ \frac{(s+\Gamma_{N})H_{1}(s)}
{(s-\mu)H_{3}(s)} \right\}(t) \right) .
\label{eq45}
\end{eqnarray}   
In the next step we subsequently calculate $\partial n_{1\uparrow}(t)/\partial\mu$ 
taking into consideration only such terms  which depend  on the bias voltage [see 
Eq.~(A2)]. As a result we get
\begin{eqnarray}
\frac{\partial n_{1\uparrow}(t)}{\partial\mu} &=&
\frac{\Gamma_{N}}{\pi} \left\{ \Gamma_{S}^{2}\lambda_{1}^{4} 
\left| {\cal{L}}^{-1} \left\{ \frac{s+\Gamma_{N}}{(s+i\mu)H_{3}(s)} \right\}(t) \right|^{2}
+ \left| {\cal{L}}^{-1} \left\{ \frac{(s+\Gamma_{N})H_{1}(s)}{(s+i\mu)H_{3}(s)} \right\}(t) \right|^{2}
\right. \nonumber \\ 
&-& \left. \Gamma_{S}^{2} \left| {\cal{L}}^{-1} \left\{ \frac{H_{1}(s)}{(s+i\mu)H_{3}(s)} \right\}(t) \right|^{2}
- \lambda_{1}^{4} \left| {\cal{L}}^{-1} \left\{ \frac{(s+\Gamma_{N})^{2}}{(s+i\mu)H_{3}(s)} \right\}(t) \right|^{2}
\right\} .
\label{eq46}
\end{eqnarray}   
\end{widetext}

The definition $G_{\uparrow}(V,t)=\frac{\partial}{\partial V} j_{N\uparrow}(V,t)$ 
along with Eqs.\ (\ref{B_s},\ref{eq45},\ref{eq46}) yield the information on time-dependent quasiparticle
states of our hybrid structure. The relevant inverse Laplace transforms can be 
obtained in a form of the linear combinations of  $\exp{\left(-\Gamma_N t\right)}$ and 
$\exp{\left( -s_{i} t\right)}$ with coefficients being 
the functions of $s_{1}, ..., s_{8}$, where $s_{1,2}=-\Gamma_N \pm i\Gamma_{S}$ and  
$s_{3}$, $s_{4}$, $s_{5}$ ($s_{6}$, $s_{7}$, $s_{8}$) are solutions of 
the cubic equation $H_{1}(s)=0$ [$H_{2}(s)=0$].

\begin{figure}
\includegraphics[width=1\columnwidth]{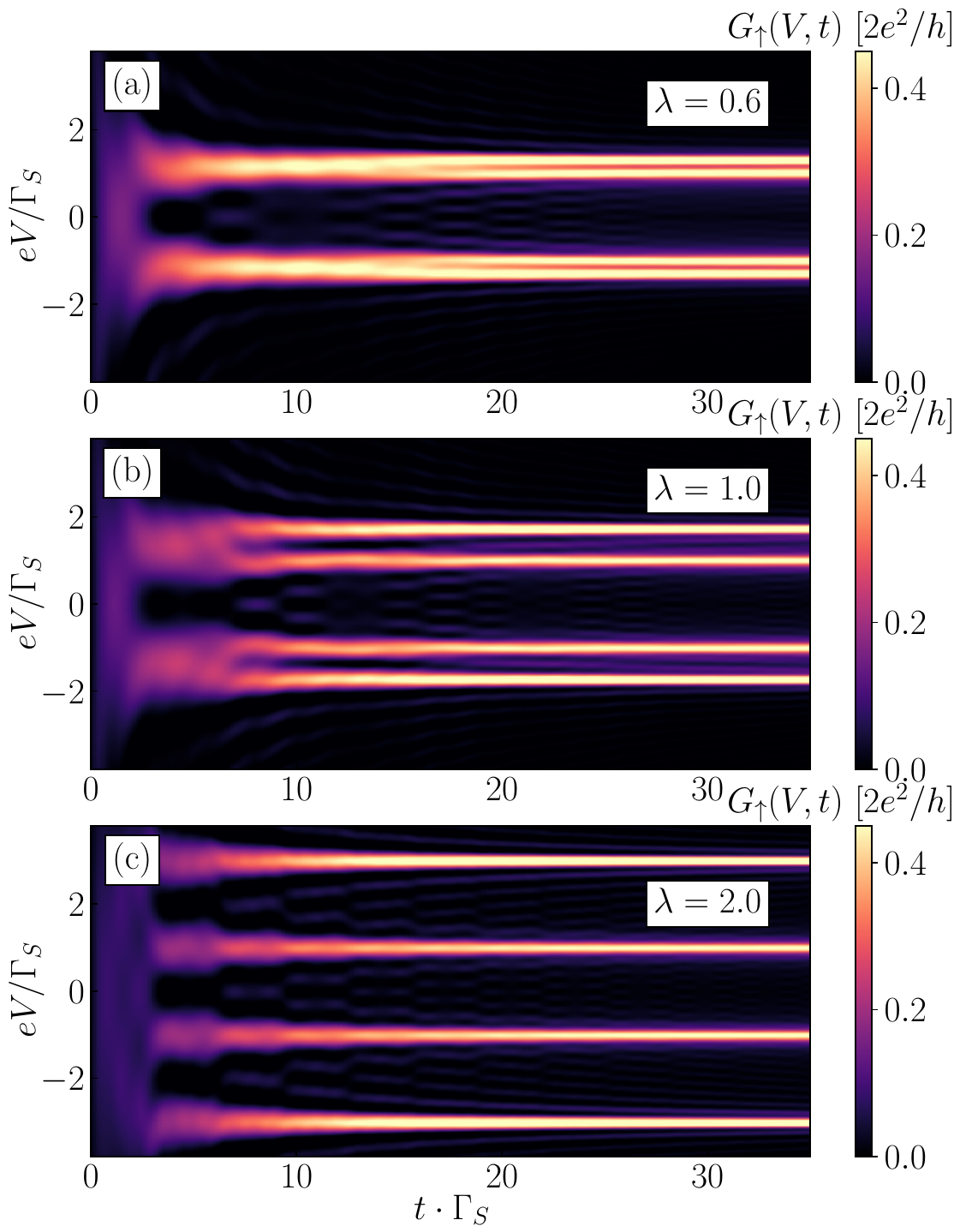}
\caption{The time-resolved differential conductance of the charge current
flowing through the first quantum dot (QD$_{1}$) induced by the voltage 
bias $V$ applied between the normal and superconducting leads.
The consecutive panels 
refer to the different couplings $\lambda\equiv\lambda_{1}=\lambda_{2}$,
as indicated. The other parameters are the same as in Fig.~\ref{Fig2} with $\Gamma_N=0.2$.}
\label{Fig5}
\end{figure}

Figure \ref{Fig5} presents the time-dependent conductance versus the bias
voltage $V$ obtained for $\lambda=0.6$, $1$ and $2$ ($\lambda\equiv\lambda_1=\lambda_2$), respectively.
It shows how the quasiparticle peaks of QD$_{1}$ develop in time. For weak 
couplings, $\lambda=0.6$ and $1$, we observe roughly two quasiparticle peaks 
emerging from the initial broad-structure, see Figs.~\ref{Fig5}(a) and (b).
Their steady-limit shape establishes at relatively long time in comparison 
to the results obtained for a stronger coupling,  $\lambda=2$. After 
a closer inspection, however, we can notice some tiny splitting between the maxima. 
In contrast, for larger coupling $\lambda$, there appear four quasiparticle 
peaks well separated from one another. Their steady limit structure establishes 
pretty fast, nearly at $t\approx 5/\Gamma_S$.
The duration of the transient region is thus 
strongly sensitive to the coupling strength $\lambda$.

\begin{figure}
\includegraphics[width=1\columnwidth]{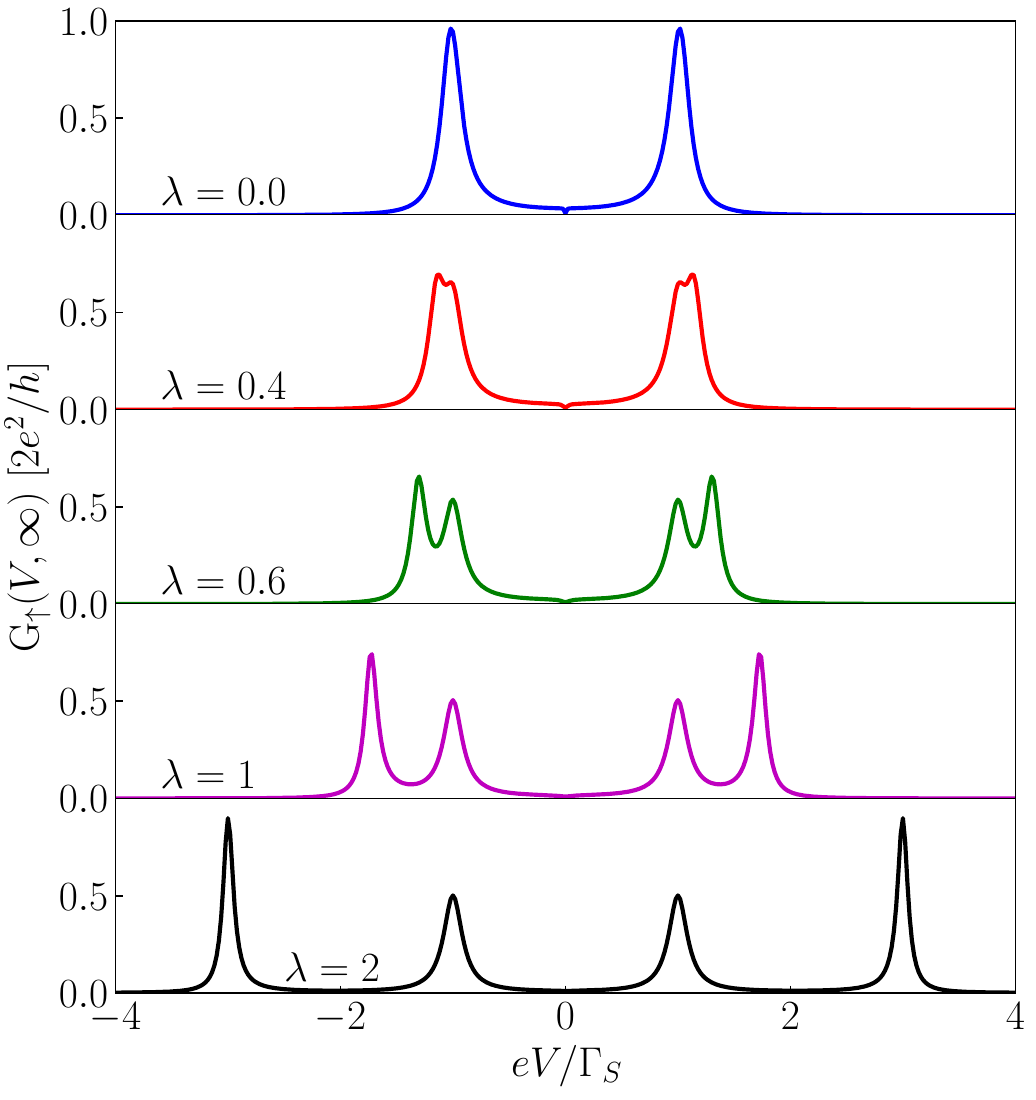}
\caption{The steady-state limit of the differential
conductance $G_{\uparrow}(V,\infty)$ 
obtained for $\varepsilon_{1}=\varepsilon_{2}=0$
and several values of $\lambda\equiv\lambda_1=\lambda_2$, 
as indicated. The other parameters are the same as in Fig.~\ref{Fig5}.}
\label{Fig6}
\end{figure}

In Fig.\ \ref{Fig6} we present the steady limit differential conductance
$G_{\sigma}(V,t\rightarrow \infty)$ obtained for different
couplings $\lambda$, as indicated.
Since one of the solutions, $s_{3}$ ($s_{6}$) for the cubic equation has 
a negative real value and the other solutions are complex (with 
negative real parts), we  get 
\begin{eqnarray}
&&G_{\uparrow}(\mu,\infty)= 2\Gamma_{N}
\mbox{\rm Re} \left( \frac{\phi_{1}(\mu)}{\phi_{3}(\mu)} \right)
-2\Gamma_{N}^{2} \left[ \left| \phi_{1}(\mu)\right|^{2} 
\right. \label{eq34} \\ 
&-& \left. \Gamma_{S}^{2} \left| \phi_{2}(\mu)\right|^{2}  + \lambda^{4}
\left( \Gamma_N^{2}+\mu^{2} \right) \left( \Gamma_{S}^{2}-\Gamma_N^{2}-\mu^{2} \right)
 \right] / \left| \phi_{3} \right|^{2} ,
\nonumber
\end{eqnarray}
where
%
$\phi_{2}(\mu) = \prod_{j=3,4,5}\left( i \mu - s_{j} \right)$, 
$\phi_{1}(\mu) = \left( i \mu + \Gamma_N \right) \phi_{2}(\mu)$, and
$\phi_{3}(\mu) = \prod_{j=1,2,6,7,8}\left( i \mu - s_{j} \right)$. 
%
We can approximately evaluate $G_{\uparrow}$, taking into consideration 
only the first term appearing in Eq.\ (\ref{eq34}). From this analysis, 
we can determine positions of peaks in the differential conductance.
We observe two peaks at positive and two 
peaks at negative bias voltage. The internal peaks at $eV = \pm \Gamma_{S}$ 
hardly depend on $\lambda$, whereas the outer peaks appear at 
$eV\simeq \pm \sqrt{\Gamma_{S}^{2}+2\lambda^{2}}$.
Moreover, the spectral weight 
of the internal peaks varies from $1$ (for $\lambda=0$) to $0.5$ 
(for stronger couplings $\lambda$). At the same time, the spectral 
weight of the outer peaks at $eV=\pm \sqrt{\Gamma_{S}^{2}+2\lambda^{2}}$
increases to $1$ (for $\lambda\geq 1$).

The time-dependent differential conductance shown Fig.\ \ref{Fig5} and 
its asymptotic limit (Fig.~\ref{Fig6}) provide information about the 
Andreev-type states of QD$_{1}$ (due to superconducting proximity effect) 
obtained for the special case, when the energy level $\varepsilon_{1}=0$ 
coincides with the zero-energy Majorana mode. Under such circumstances, 
the influence of the Majorana mode on the quasiparticle spectrum of QD$_{1}$ 
is manifested by destructive quantum interference in  analogy to
Fano-type lineshapes appearing in double quantum dot T-shape configurations
\cite{Baranski-2019,Gorski-2018,Baranski-2017}. 
Here, QD$_{1}$ is at the interface between the superconducting
and normal leads with the side-attached Majorana mode,
playing the role of ``second quantum dot''. Electrons moving between
external leads can hop aside to the zero-energy level
of the nanowire and return with a different phase, giving rise to destructive quantum interference, thereby depleting the spectral function of QD$_1$ near $\omega=\varepsilon_{1}$. Such situation is no longer present for $\varepsilon_{1}\neq 0$ because the side-attached Majorana mode has nothing to interfere with.
In the present case, this destructive interference shows up as a tiny dip 
at zero voltage for a weakly coupled heterostructure (see the curves
presented  in Fig.\ \ref{Fig6} for $\lambda=0.4$ and $\lambda=0.6$).

In the remaining part of this section, we investigate the quasiparticle 
features appearing in the conductance $G_{\uparrow}(\mu,t)$ for 
$\varepsilon_{1}\neq 0$, while $\varepsilon_2 = 0$.
In such a situation the Majorana mode 
has a constructive influence on the transport properties,
inducing the zero-energy quasiparticle 
which enhances the zero-bias conductance. This effect resembles 
a typical Majorana leakage on the quantum dot hybridized to 
the normal (metallic) electrodes \cite{Vernek-2015,Prada-2017,Klinovaja-2017,
Ptok-2017,Weymann-2017,Deng-2018,Silva-2020,Seridonio-2020,Majek_Weyman-2021,
Domanski-2023,Vernek-2023,Cayao_2023}. For nonzero $\varepsilon_{1}$,
we obtain 
\begin{eqnarray}
G_{\uparrow}(\mu,t)&=&2\Gamma_{N} \mbox{\rm Re}\left[ e^{-i\mu t} 
{\cal{L}}^{-1} \left\{ \frac{(s+i\varepsilon +\Gamma_{N})F_{1}(s)}{(s-i\mu)F_{2}(s)} \right\}(t)
 \right] \nonumber \\ 
&-&2\Gamma_{N}^{2} \left[ \left| 
{\cal{L}}^{-1} \left\{ \frac{(s+i\varepsilon +\Gamma_{N})F_{1}(s)}{(s-i\mu)F_{2}(s)} \right\}(t)
\right|^{2} \right. \nonumber \\
&+&\Gamma_{S}^{2} \lambda_{1}^{4} \left| 
{\cal{L}}^{-1} \left\{ \frac{s+i\varepsilon +\Gamma_{N}}{(s-i\mu)F_{2}(s)} \right\}(t)
\right|^{2} \nonumber \\
&-& \Gamma_{S}^{2} \left| 
{\cal{L}}^{-1} \left\{ \frac{F_{1}(s)}{(s+i\mu)F_{2}(s)} \right\}(t)
\right|^{2}  \nonumber \\
&-&\lambda_{1}^{4} \left. \left| 
{\cal{L}}^{-1} \left\{ \frac{(s+\Gamma_{N})^{2}+\varepsilon^{2}}{(s+i\mu)F_{2}(s)} \right\}(t)
\right|^{2}  \right] ,
\end{eqnarray}
where $F_{1}(s)=s^{3}+2\Gamma_N s^{2}+s(\varepsilon^{2}+\Gamma_{S}^{2}+\Gamma_{N}^{2}+\lambda_{1}^{2})
-\lambda_{1}^{2}(i\varepsilon-\Gamma_N)$ and $F_{2}(s)=(s^{2}+2\Gamma_N s+\varepsilon^{2}+
\Gamma_{S}^{2}+\Gamma_{N}^{2})(s^{3}+2\Gamma_{N}s^{2}+s(\Gamma_N^{2}+\varepsilon^{2}+\Gamma_{S}^{2}+2\lambda_{1}^{2})
+2\lambda_{1}^{2}\Gamma_N)$.

\begin{figure}
\includegraphics[width=1\columnwidth]{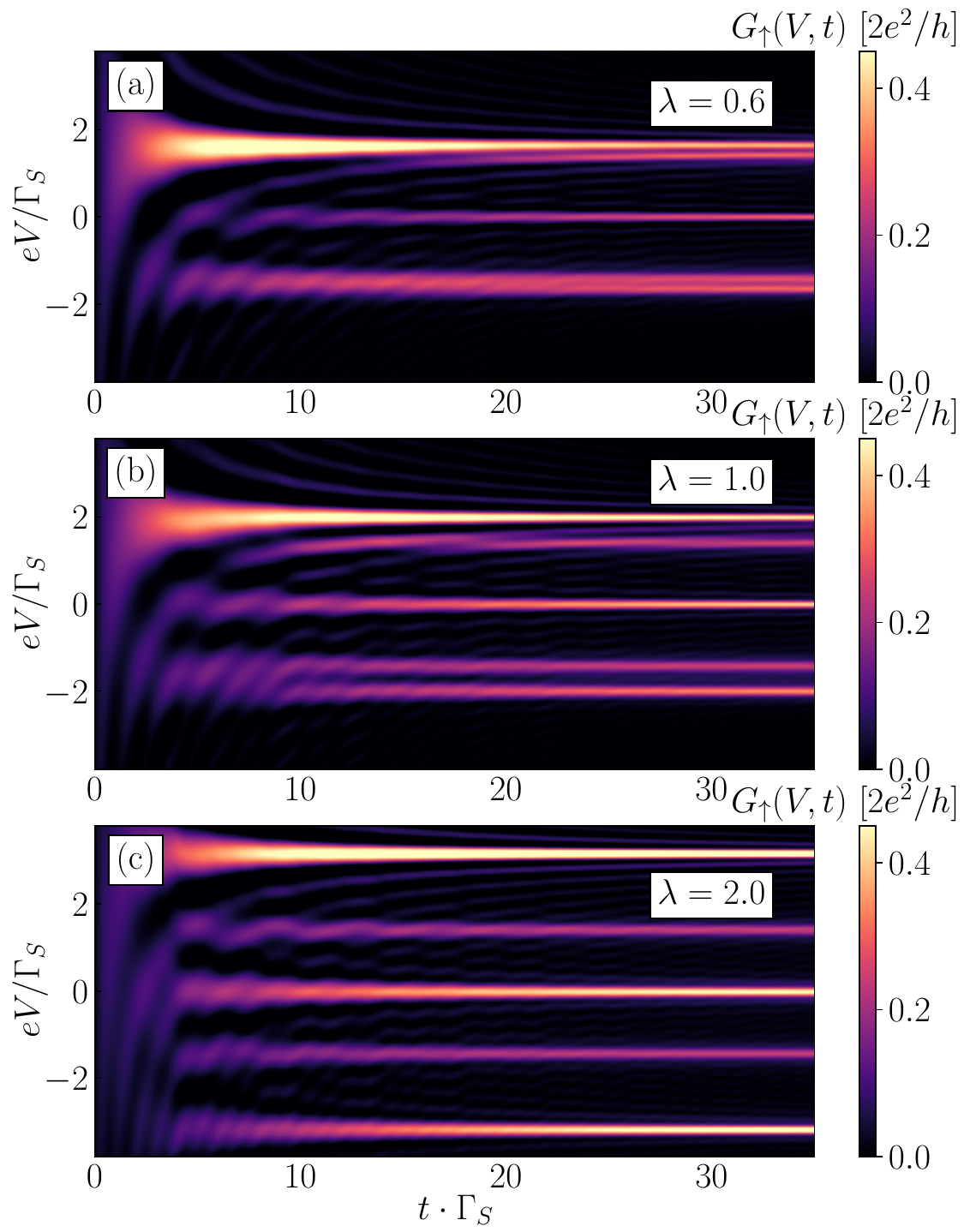}
\caption{The time-dependent differential conductance $G_{\uparrow}(V,t)$ obtained 
for non-zero energy level of the first quantum dot, $\varepsilon_{1}=2$, 
while $\varepsilon_2=0$, and 
for different values of the coupling $\lambda$, as indicated.
The other parameters are the same as in Fig.~\ref{Fig5}.}
\label{Fig7}
\end{figure}

\begin{figure}
\includegraphics[width=1\columnwidth]{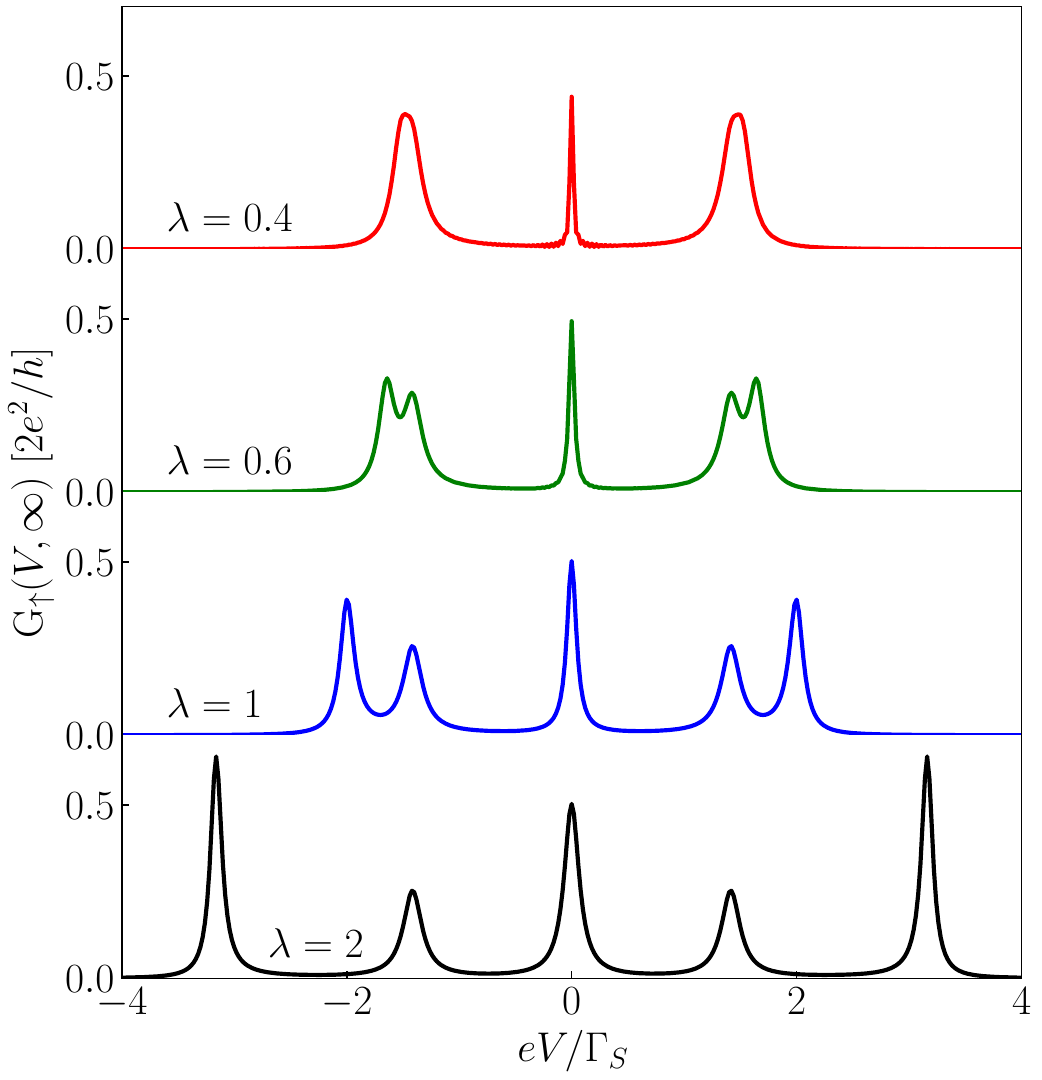}
\caption{The steady-state limit of the differential
conductance $G_{\uparrow}(V,t=\infty)$ 
obtained for several values of $\lambda$ (as indicated) 
and the other parameters the same as in Fig.~\ref{Fig7}.}
\label{Fig8}
\end{figure}

Figure \ref{Fig7} shows the time-dependent differential conductance
$G_{1\uparrow}(V,t)$ obtained for non-zero value of the QD$_{1}$ energy level, 
$\varepsilon_{1}=2$, and several values of the coupling $\lambda$, as indicated.
In the stationary limit ($t \rightarrow \infty$), the height of the zero-energy 
peak tends to $0.5$, which is a typical fractional value initially predicted for 
the leaking Majorana mode \cite{Baranger-2011,Vernek-2015,Prada-2017} 
for arbitrary $\lambda$. Its width increases here upon increasing 
the coupling $\lambda$ as can be also seen in the stationary limit 
presented in Fig.\ \ref{Fig8}, which displays the behavior of the differential conductance in the stationary limit.

\begin{figure}
\includegraphics[width=1\columnwidth]{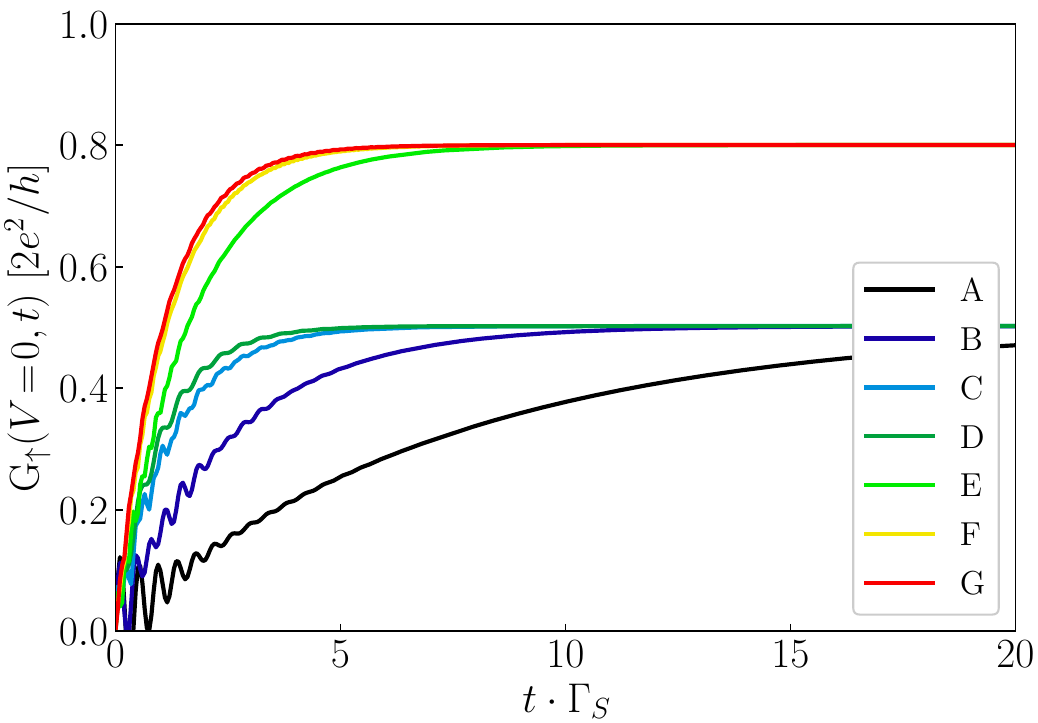}
\caption{The zero-bias differential conductance $G_{\uparrow}(V=0,t)$ 
varying against time for $\varepsilon_{1}=1$ (A, B, C, D curves) 
and $\varepsilon_{1}=2$ (E, F, G lines) using different values of
$\lambda$ listed in Tab.~\ref{tab1}.
The other parameters are the same as in Fig.~\ref{Fig7}.
}
\label{Fig9}
\end{figure}

To estimate the time interval in which the zero-energy Majorana mode 
leaks onto QD$_{1}$ we introduce a phenomenological parameter
$\tau$ defined by \cite{Baranski-2021}
\begin{equation}
G_{\uparrow}(0,t) = G_{\uparrow}(0,\infty) \left[ 1 - e^{-t/t} \right] .
\end{equation}
This parameter characterizes the time-scale, over which the  zero-bias 
conductance approaches the stationary limit value $ G_{\uparrow}(0,\infty)$. 
We computed $\tau$ for two different values of energy levels $\varepsilon_{1}$ 
(while $\varepsilon_2=0$) and 
several values of the coupling $\lambda$. Figure\ \ref{Fig9} presents 
the zero-bias conductance $G_{\uparrow}(V=0,t)$ for $\varepsilon_1=1$ 
(A, B, C, D curves) and $\varepsilon_1=2$ 
(E, F, G lines), using $\lambda$ specified 
in Tab.~\ref{tab1}. We found $\tau \approx 7.5$, $2.5$, $1.25$, $1.0$ 
(expressed in units of $\frac{1}{\Gamma_{N}}$) for the couplings 
$\lambda=0.4$, $0.8$, $2$ and $8$, respectively.
Similarly, for the other value of the 
energy level, $\varepsilon_{1}=2$, we estimated the leakage time-scales 
$\tau \approx 1.6$, $1.15$, $1.1$. We thus conclude that the development 
of the zero-energy Majorana mode on QD$_{1}$ occurs faster upon increasing 
the coupling to topological superconductor $\lambda$.

\begin{table}[h!]
  \begin{center}
    \begin{tabular}{|c|c|c|c|}
\hline
      \text{Curve} & $\lambda [\Gamma_{S}]$ & 
      $\varepsilon_{1}  [\Gamma_{S}]$ & $\tau [1/\Gamma_{N}]$\\
      \hline
A & 0.4  & 1 & 7.5 \\
B & 0.8  & 1 & 2.5 \\
C & 2.0  & 1 & 1.25 \\
D & 8.0  & 1 & 1.0 \\
E & 2.0  & 2 & 1.6 \\
F & 4.0  & 2 & 1.15 \\
G & 6.0  & 2 & 1.1 \\
\hline
    \end{tabular}
    \caption{The time-scale $\tau$ of the Majorana leakage obtained for several 
couplings $\lambda\equiv\lambda_{1}=\lambda_{2}$ and different values of the first quantum dot energy level $\varepsilon_{1}$ with $\varepsilon_{2}=0$ }
    \label{tab1}
  \end{center}
\end{table}

Summarizing this section, we emphasize that quasiparticles emerging 
on QD$_{1}$ (presented in Figs.~\ref{Fig7} and \ref{Fig8}) represent 
the trivial Andreev bound states (at finite voltage $V$) coexisting 
with the topological (zero-bias) feature. Their formation occurs over some 
characteristic time-scale $\tau$ [see Table \ref{tab1}] and is accompanied
by the damped quantum oscillations. Buildup of the quasiparticles  
is predominantly controlled by the coupling of QD$_{1}$ to a continuous 
spectrum of the metallic lead, but spectral weights and energies of such 
quasiparticles depend on the energy level $\varepsilon_{1}$ of 
QD$_{1}$, what indirectly affects the profile of damped quantum 
oscillations observed in the transient regime. 

Let us also comment about the characteristic energy and time scales of our setup which could be verified empirically.
Topological superconductivity of the semiconducting nanowires and magnetic nanochains has been so far achieved by contacting them with the conventional superconductors, such as Al or Pb whose pairing gaps (safely below $T_{c}$) are about 0.5 meV. The topological gap (which separates the zero-energy Majorana modes from the trivial bound states) is even smaller, on the order of 0.1-0.2 meV. This establishes the typical energy scale for our hybrid system, because within the effective low-energy description (\ref{M-DQD}) we consider only the Majorana modes of nanowire, neglecting any other (higher-energy) trivial bound states. As far as the coupling $\Gamma_{S}$ of QD$_1$ is concerned, we assume it to be at most comparable to the topological gap of the nanowire (because otherwise QD1 would hybridize with trivial states of the nanowire). The other coupling $\Gamma_{N}$= 0.1$\Gamma_{S}$, controlling the relaxation processes, is on the order of $~$10 $\mu$eV. Under such conditions the typical transient region would extend from a fraction to a few nanoseconds [for more detailed quantitative evaluations for the single dot – topological superconductor hybrid structure see Ref. \cite{Baranski-2021}).

\section{Correlation effects}
\label{correlations}

Finally, we address the role played by the Coulomb repulsion
$U\hat{n}_{i\uparrow}\hat{n}_{i\downarrow}$, which can be expected to 
suppress the superconducting proximity effect. We shall study its influence 
both on the local and nonlocal electron pairings. For a single quantum 
dot attached to superconductor, this issue has been extensively studied, 
considering the static \cite{Bauer-2007,Rodero-2011} and nonequilibrium 
conditions \cite{LevyYeyati-2021,Morr-2022,Domanski_etal_2023,Wegewijs_2023}.
Depending on the ratio of $\Gamma_{S}/U$ and the energy level, the quantum dot was predicted to be either singly occupied or in the BCS-type configuration.
The superconducting proximity effect is efficient only in the latter case.   
In this section we analyze effects of the Coulomb repulsion 
on the time-dependent pairings in the transient region and in 
asymptotic behavior of our hybrid structure (Fig.~\ref{scheme}).

To accurately describe the correlation effects and system's dynamics,
we resort to the numerical renormalization group (NRG) method ~\cite{Wilson1975,Bulla2008,NRG_code}.
This method has been successfully used to analyze the stationary properties of the Anderson impurity coupled to superconductor \cite{Bauer-2007,Domanski-2016}. Here, we make use of its time-dependent extension \cite{Anders2005,Costi2014,WrzesniewskiWeymann-2019} to address the dynamical effects of our setup. The main idea of the NRG approach is a logarithmic discretization of the conduction band, which allows one to map the Hamiltonian to a chain-like form. Matrix Hamiltonian of such a model can be next diagonalized in an iterative fashion, keeping an appropriate number of the low-energy states. This technique can be used to determine the time-dependent physical observables and is not limited by perturbative approximations (for details see e.g. Refs.~\cite{Anders2005,Costi2014,WrzesniewskiWeymann-2019}).

\begin{figure}[b!]
\includegraphics[width=1\columnwidth]{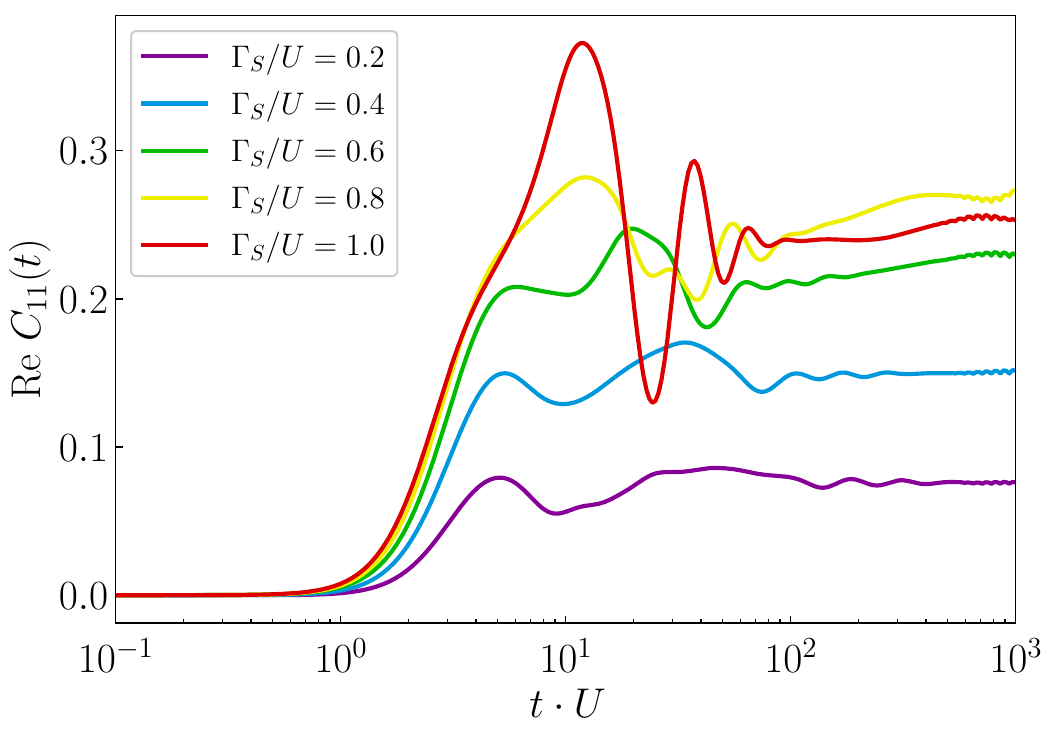}
\caption{The real part of $C_{11}(t)=\langle d_{1\downarrow}(t) d_{1\uparrow}(t)\rangle$ obtained by time-dependent NRG calculations for the half-filled quantum dots 
$\varepsilon_{1}=\varepsilon_{2}=-U/2$, assuming $\Gamma_{N}/U=0.1$ and $\lambda/U=0.2$ ($\lambda\equiv\lambda_1=\lambda_2$), with $U=1$, and different values of $\Gamma_{S}$, as indicated.
Energies are now expressed in terms of band halfwidth.
}
\label{Fig_NRG1}
\end{figure}

In Fig.~\ref{Fig_NRG1} we present the time-dependent local electron 
pairings $C_{11}(t)=\langle d_{1\downarrow}(t) d_{1\uparrow}(t)\rangle$
obtained for the half-filled quantum dots $\varepsilon_{i}=-U/2$ and different values of $\Gamma_{S}/U$. In this figure the quench is performed in all the couplings, i.e. the couplings to both leads and topological superconductor.
To identify temporal extent of the transient region, we plot all observables against the logarithm of time. 
First of all, one can notice that transient phenomena start developing when $t\gtrsim 1/U$ and are most important for time scales coinciding roughly with $t\sim 1/\Gamma_{N}$.
Moreover, a closer inspection of the local electron pairing
induced on QD$_{1}$ reveals its substantial
suppression caused by the Coulomb repulsion.
This tendency is consistent with the steady-state 
solution for N-QD$_{1}$-S nanostructure \cite{Bauer-2007,Domanski-2016}. 
The reduction of the local electron pairing is associated with ensuing changeover 
of the quantum dot ground state, from the BCS-type to the singly occupied 
configuration. In the absence of the topological superconductor ($\lambda=0$)
such quantum phase transition would occur at $\Gamma_{S}/U=0.5$ \cite{Bauer-2007}.
However, the influence of the side-attached topological superconductor partly smears this 
singlet-doublet phase transition in our setup \cite{Majek_Weyman-2021,Domanski-2023}.

\begin{figure}
\includegraphics[width=1\columnwidth]{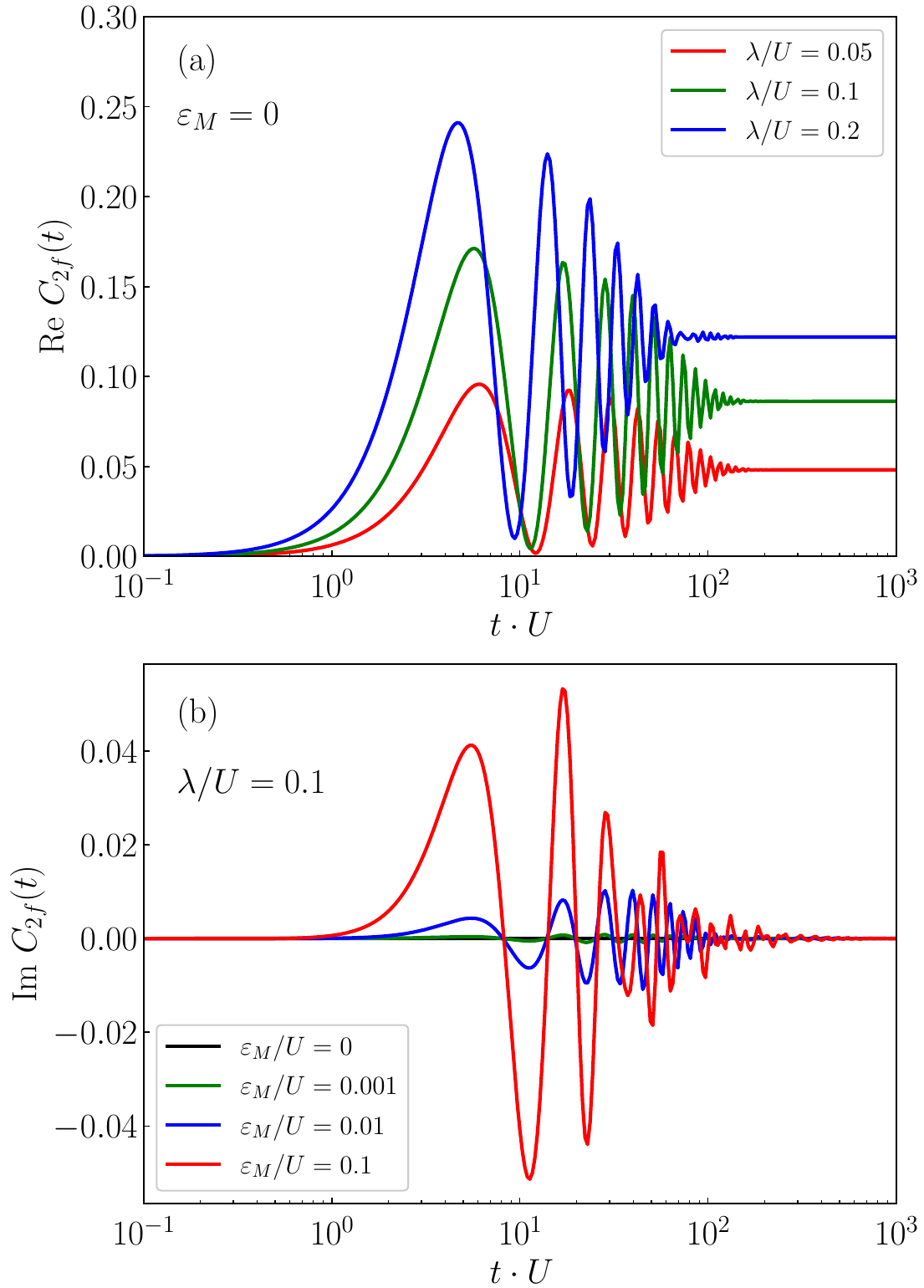}
\caption{The variation of the (a) real and (b) imaginary parts of $C_{2f}(t)$ obtained 
by time-dependent NRG calculations for the half-filled quantum dots 
$\varepsilon_{1}=\varepsilon_{2}=-U/2$, assuming 
$\Gamma_{N}/U=0.1$, coupling $\Gamma_{S}/U=0.75$
and different values of (a) $\lambda\equiv\lambda_1=\lambda_2$
and (b) the overlap between the Majorana modes $\varepsilon_M$, as indicated.}
\label{Fig_NRG2}
\end{figure}

Let us now focus on the dynamics of the triplet pairing $C_{2f}(t) = \expect{ \hat{d}_{2\uparrow}(t) \hat{f}(t) }$,
which is presented in Fig.~\ref{Fig_NRG2}. To analyze its behavior, we fix the couplings to the normal and superconducting contacts and perform quench only in the coupling to topological superconductor. The upper panel presents 
the evolution of the real part of $C_{2f}(t)$ for different values of the couplings to topological superconductor in the case of $\varepsilon_M=0$, while the bottom panel shows the imaginary part of $C_{2f}(t)$ for one selected value of $\lambda$,
while assuming finite $\varepsilon_M$, as indicated.
We first notice that the mixed (triplet) pairing,
in contrast to the local pairing induced at QD$_{1}$,
hardly depends on $\Gamma_S$. This property indicates that the leakage of the Majorana mode on the side-attached quantum dots is not endangered by the interplay of correlations and superconducting pairing on the first quantum dot.
Finite $C_{2f}(t)$ starts developing when $t\gtrsim 1/U$, showing considerable oscillations for $t\gtrsim 1/\Gamma_N$, which however die out at longer times. As can be seen, the amplitude of these oscillations increases with raising coupling to topological superconductor $\lambda$, giving rise to larger value of $C_{2f}(t)$ in the long time limit, see Fig.~\ref{Fig_NRG2}(c).
On the other hand, the imaginary part of $C_{2f}(t)$ shows considerable oscillations in the transient regime only when there is a finite overlap between Majorana modes
(the real part hardly depends on $\varepsilon_M$), see Fig.~\ref{Fig_NRG2}(b).
The amplitude of these oscillations increases with $\varepsilon_M$, indicating a non-local character of Majorana quasiparticles in the transient regime.

\section{Summary and outlook}
\label{summary}

We have studied the local and nonlocal transient phenomena of
a hybrid structure composed of two quantum dots attached on opposite 
sides to the topological superconducting nanowire. We have shown that 
in a steady-state limit these spatially distant quantum dots interconnected 
through the Majorana edge-modes develop the quasiparticle spectra 
independent of one another. Despite the lack of static cross-correlations,
however, we have found dynamical nonlocal effects transmitted between the dots
surviving over a finite time interval.
These effects are apparent in the inter-dot electron pairing,
both for the singlet and triplet channels, which could be 
detectable in the crossed Andreev reflection spectroscopy.

We have also investigated gradual development of the quasiparticle 
states of QD$_{1}$ (placed) on interface between the normal and 
conventional superconductor), focusing on signatures of the Majorana mode. We 
have found coexistence of the Andreev (trivial) bound states
with the  zero-energy (topological) feature. Emergence of these 
quasiparticles can be empirically verified by time-resolved 
differential conductance of the tunneling current induced across QD$_{1}$ 
by the bias voltage between the external metallic 
and superconducting leads. For the zero energy level $\varepsilon_{1}=0$, 
the Majorana mode imprints a tiny dip at zero voltage, originating from
its destructive quantum interference on the spectrum of QD$_{1}$. 
This effect is analogous to the Fano-type interference  
observed in T-shape configurations of double quantum dot 
junctions \cite{Baranski-2019,Gorski-2018,Baranski-2017}. 
Otherwise, i.e.\ for $\varepsilon_{1}\neq 0$, the Majorana 
mode induces the zero-energy conductance peak. In the latter 
case the leakage of Majorana mode yields a fractional value of
the zero-bias differential conductance, Fig.\ \ref{Fig8}.  
We have also studied the effect of Coulomb interactions 
on the transient phenomena and revealed their destructive 
influence on the local (on-dot) pairing.

Our analysis relied on the assumption of very large pairing 
gap $\Delta_{SC}$ of the superconducting lead. In realistic 
situations one should additionally take into account the electronic
states from outside this pairing gap, which for conventional 
superconductors is typically a fraction of meV. The influence 
of such continuum has been discussed for superconducting 
nanostructures, using various many-body techniques  
\cite{Souto_2016,LevyYeyati-2017,Eckstein-2021,Sothmann-2021,
LevyYeyati-2021,Morr-2022,Wegewijs_2023} suitable to capture 
also the correlation effects. These methods could be adopted 
to the present setup as well. In general,  the quasiparticles
from outside the pairing gap would contribute to the relaxation
processes, partly reducing the relevant time scales characterizing 
the transient phenomena. Another important issue could be associated 
with qualitative changeover of the trivial (Andreev) and 
topological (Majorana) states imposed by the quantum 
quenches \cite{Tuovinen-2021,Dagotto-2023}. Post-quench evolution
could lead to the dynamical phase transition \cite{Domanski_etal_2023},
but this challenging topic is beyond the scope of the present study.

For possible verification, we have evaluated the time-scale 
needed for emergence of the trivial (Andreev) and topological (Majorana) 
bound states. In typical realizations of superconducting hybrid structures  
(where the couplings $\Gamma_{N,S}$ of the quantum dots to external leads 
are $\sim$ meV) the duration of the transient effects would cover nanoseconds 
region. Currently available tunneling spectroscopies \cite{review_time_resolved} should be 
able to detect the nonlocal cross-correlations between the spatially 
distant quantum dots embedded into the setup presented in Fig.~\ref{scheme}. 
Detection of transient nonlocal pairing
in the singlet $C_{12}(t)$ and triplet $C_{12}^{\rm T}(t)$ channel could be
obtained by crossed Andreev spectroscopy, analogous to
the methods used recently in the minimal-length topological superconducting
system \cite{Kouwenhoven-2023,Bordin2024Feb}. As concerns the triplet channel, 
its measurement would be feasible  by means of the equal spin 
electron-to-hole scattering \cite{He_2014}. Time-resolved nonlocal 
spectroscopy could be done for the hybrid structure, using either 
semiconducting nanowires (e.g. InAs) partly covered by conventional 
superconductors (like in the setup reported in Ref. \cite{Deng-2016}) 
or by depositing magnetic atoms (Fe) on superconductors and attaching other non-magnetic impurities to them that can be probed by scanning
spin-polarized Andreev spectroscopy \cite{Sun_2016}.
Detailed knowledge of such transient nonlocal 
effects might be useful for reliable control of braiding
protocols for conventional and/or topological
superconducting quantum bits \cite{Flensberg_2020,Mascot_2023}.

Finally, we would like to mention that 
it would be worthwhile to extend the present analysis
of the nonlocal transient phenomena onto hybrid structures with 
quasi-Majorana states. These quasiparticles have been predicted
to form at the boundaries of semiconducting nanowires proximitized
to superconducting materials, due to a flat confining
potential \cite{Kells-2012}, a suppressed superconducting 
pair potential and/or an excess Zeeman field \cite{Roy-2013} or 
attachment of quantum dot(s)
\cite{DasSarma-2017}. Under such circumstances the trivial Andreev 
states appear at nearly zero-energy, and their properties,
such as e.g. the zero-bias conductance peak \cite{Moore-2018,Setiawan-2017},
the fractional Josephson effect \cite{Chiu-2019},
and the braiding schemes \cite{Wimmer-2019},  
very much resemble the behavior of true Majorana modes.
In static situations, it is rather
difficult to discern whether the zero-energy edge modes have their trivial or topological origin \cite{Tewari-2018,Liu-2018}.
However, the theoretical results reported in Ref.\ \cite{Tuovinen_2019} seem to indicate 
that the true Majorana modes could be unambiguously manifested 
by the characteristic quantum oscillations in time-dependent conductance right after applying a source-drain voltage, which otherwise would not occur for the trivial modes.
In this regard, a follow-up in-depth study of the
nonlocal transient effects for quasi-Majorana modes
in similar hybrid devices as considered here would be desirable.

\begin{acknowledgments}
This work was supported by the National Science Centre
in Poland through the Project No. 2018/29/B/ST3/00937.
K.W. acknowledges financial support by the National Science Centre
in Poland through the Project No. 2022/45/B/ST3/02826.
We acknowledge the computing time at the Pozna\'n Supercomputing and Networking Center.
We thank T. Kwapi\'nski and B. Baran for technical assistance.
\end{acknowledgments}

\appendix


\section{Time-dependent expectation values}
\label{app}

To calculate the electron occupancy of QD$_{1}$ embedded in the uncorrelated 
setup we use the expression (\ref{d1up}) for $\hat{d}_{1\uparrow}(s)$ 
and obtain
\onecolumngrid
\begin{eqnarray}
n_{1\uparrow}(t) &=& \left< {\cal{L}}^{-1} \left\{\hat{d}_{1\uparrow}^{\dagger}(s)
\right\}{\cal{L}}^{-1} \left\{\hat{d}_{1\uparrow}(s)\right\} \right>
\label{A1} \\ &=& 
\left[ 1 - n_{1\uparrow}(0) \right] \lambda_{1}^{4} \left(
{\cal{L}}^{-1} \left\{ \frac{(s+\Gamma_{N})^{2}}{H_{3}(s)} \right\} (t) \right)^{2}
+  n_{1\uparrow}(0)  \left( {\cal{L}}^{-1} \left\{ \frac{(s+\Gamma_{N})H_{1}(s)}
{H_{3}(s)} \right\} (t) \right)^{2}
\nonumber\\ & + &
\left[ 1 - n_{1\downarrow}(0) \right] \Gamma_{S}^{2} \left(
{\cal{L}}^{-1} \left\{ \frac{H_{1}(s)}{H_{3}(s)} \right\} (t) \right)^{2}
+  n_{1\downarrow}(0) \Gamma_{S}^{2}\lambda_{1}^{4} \left(
{\cal{L}}^{-1} \left\{ \frac{s+\Gamma_{N}}{H_{3}(s)} \right\} (t) \right)^{2}
\nonumber \\ & + &
\expect{ \hat{f}^{\dagger}(0)+\hat{f}(0)}
\expect{ \hat{f}(0)+\hat{f}^{\dagger}(0)} \frac{\lambda_{1}^{2}}{2}
\left( {\cal{L}}^{-1} \left\{ \frac{(s+\Gamma_{N})}{H_{2}(s)} \right\}(t) \right)^{2}
+ \sum_{{\bf k},{\bf k}_{1}} V_{\bf k} V_{{\bf k}_{1}}
\left< {\cal{L}}^{-1}\left\{ S_{\bf k}^{\dagger}(s)\right\}(t) 
{\cal{L}}^{-1}\left\{ S_{{\bf k}_{1}}(s)\right\}(t) \right>,
\nonumber
\end{eqnarray}   
where $H_{1}(s)$, $H_{2}(s)$, $H_{3}(s)$ and $S_{\bf k}(s)$ are defined in Eqs. (\ref{H1},\ref{H2},\ref{H3},\ref{S}).
The first five terms can be rewritten in the form given by Eq. (\ref{n_1up})
and calculation of the last term can be performed as follows
\begin{eqnarray}
&&\sum_{{\bf k},{\bf k}_{1}} V_{\bf k} V_{{\bf k}_{1}}
\left< {\cal{L}}^{-1}\left\{ \Gamma_{S} \frac{\hat{c}_{{\bf k}\downarrow}
(0)H_{1}(s)}{(s+i\varepsilon_{\bf k})H_{3}(s)} + i \frac{\hat{c}_{{\bf k}\uparrow}^{\dagger}
(0)(s+\Gamma_{N})H_{1}(s)}{(s-i\varepsilon_{\bf k})H_{3}(s)} + \Gamma_{S}\lambda_{1}^{2}
\frac{\hat{c}_{{\bf k}\downarrow}^{\dagger}(0)(s+\Gamma_{N})}{(s-i\varepsilon_{\bf k})H_{3}(s)}
-i\lambda_{1}^{2} \frac{\hat{c}_{{\bf k}\uparrow}(0)(s+\Gamma_{N})^{2}}{(s+i\varepsilon_{\bf k})H_{3}(s)}
\right\}(t) 
\right. \nonumber \\  & \times &\left. {\cal{L}}^{-1}\left\{ 
\Gamma_{S} \frac{\hat{c}^{\dagger}_{{\bf k}_{1}\downarrow}
(0)H_{1}(s)}{(s-i\varepsilon_{{\bf k}_{1}})H_{3}(s)} - i \frac{\hat{c}_{{\bf k}_{1}\uparrow}
(0)(s+\Gamma_{N})H_{1}(s)}{(s+i\varepsilon_{{\bf k}_{1}})H_{3}(s)} + \Gamma_{S}\lambda_{1}^{2}
\frac{\hat{c}_{{\bf k}_{1}\downarrow}(0)(s+\Gamma_{N})}{(s+i\varepsilon_{{\bf k}_{1}})H_{3}(s)}
+i\lambda_{1}^{2} \frac{\hat{c}_{{\bf k}_{1}\uparrow}^{\dagger}(0)(s+\Gamma_{N})^{2}}{(s-i\varepsilon_{{\bf k}_{1}})H_{3}(s)}
\right\}(t) \right> 
\\ &=& \sum_{{\bf k},{\bf k}_{1}} V_{\bf k} V_{{\bf k}_{1}} \left\{ \Gamma_{S}^{2} 
\left< \hat{c}_{{\bf k}\downarrow}(0) \hat{c}_{{\bf k}_{1}\downarrow}^{\dagger}(0) \right>
{\cal{L}}^{-1} \left\{ \frac{H_{1}(s)}{(s+i\varepsilon_{\bf k})H_{3}(s)}\right\}(t) 
{\cal{L}}^{-1} \left\{ \frac{H_{1}(s)}{(s-i\varepsilon_{{\bf k}_{1}})H_{3}(s)}\right\}(t)
\right. \nonumber \\ &+&
\left< \hat{c}_{{\bf k}\uparrow}^{\dagger}(0) \hat{c}_{{\bf k}_{1}\uparrow}(0) \right>
{\cal{L}}^{-1} \left\{ \frac{(s+\Gamma_{N})H_{1}(s)}{(s-i\varepsilon_{\bf k})H_{3}(s)}\right\}(t) 
{\cal{L}}^{-1} \left\{ \frac{(s+\Gamma_{N})H_{1}(s)}{(s+i\varepsilon_{{\bf k}_{1}})H_{3}(s)}\right\}(t)
\nonumber \\ &+& \Gamma_{S}^{2}\lambda_{1}^{4}
\left< \hat{c}_{{\bf k}\downarrow}^{\dagger}(0) \hat{c}_{{\bf k}_{1}\downarrow}(0) \right>
{\cal{L}}^{-1} \left\{ \frac{(s+\Gamma_{N})}{(s-i\varepsilon_{\bf k})H_{3}(s)}\right\}(t) 
{\cal{L}}^{-1} \left\{ \frac{(s+\Gamma_{N})}{(s+i\varepsilon_{{\bf k}_{1}})H_{3}(s)}\right\}(t)
\nonumber \\ &+& \left. \lambda_{1}^{4}
\left< \hat{c}_{{\bf k}\uparrow}(0) \hat{c}_{{\bf k}_{1}\uparrow}^{\dagger}(0) \right>
{\cal{L}}^{-1} \left\{ \frac{(s+\Gamma_{N})^{2}}{(s+i\varepsilon_{\bf k})H_{3}(s)}\right\}(t) 
{\cal{L}}^{-1} \left\{ \frac{(s+\Gamma_{N})^{2}}{(s-i\varepsilon_{{\bf k}_{1}})H_{3}(s)}\right\}(t)
 \right\} \nonumber
\\ &=& \frac{\Gamma_{N}}{\pi} \int_{-\infty}^{\infty} d\varepsilon
\left\{ \Gamma_{S}^{2} 
\left[ 1 - f_{N}(\varepsilon)\right]
{\cal{L}}^{-1} \left\{ \frac{H_{1}(s)}{(s+i\varepsilon)H_{3}(s)}\right\}(t) 
{\cal{L}}^{-1} \left\{ \frac{H_{1}(s)}{(s-i\varepsilon)H_{3}(s)}\right\}(t)
\right. \nonumber \\ &+&
f_{N}(\varepsilon)
{\cal{L}}^{-1} \left\{ \frac{(s+\Gamma_{N})H_{1}(s)}{(s-i\varepsilon)H_{3}(s)}\right\}(t) 
{\cal{L}}^{-1} \left\{ \frac{(s+\Gamma_{N})H_{1}(s)}{(s+i\varepsilon)H_{3}(s)}\right\}(t)
\nonumber \\ &+& \Gamma_{S}^{2}\lambda_{1}^{4}
f_{N}(\varepsilon)
{\cal{L}}^{-1} \left\{ \frac{(s+\Gamma_{N})}{(s-i\varepsilon)H_{3}(s)}\right\}(t) 
{\cal{L}}^{-1} \left\{ \frac{(s+\Gamma_{N})}{(s+i\varepsilon)H_{3}(s)}\right\}(t)
\nonumber \\ &+& \left. \lambda_{1}^{4}
\left[ 1 - f_{N}(\varepsilon)\right]
{\cal{L}}^{-1} \left\{ \frac{(s+\Gamma_{N})^{2}}{(s+i\varepsilon)H_{3}(s)}\right\}(t) 
{\cal{L}}^{-1} \left\{ \frac{(s+\Gamma_{N})^{2}}{(s-i\varepsilon)H_{3}(s)}\right\}(t)
 \right\} . \nonumber
\end{eqnarray}   
This expression can be simply transformed to the form given in Eq.\ (\ref{n_1up}).

Performing similar calculations for $n_{1\downarrow}(t)$ we obtain  
\begin{eqnarray}
n_{1\downarrow}(t) = n_{1\uparrow}(0) N_{\uparrow}(t)+
n_{1\downarrow}(0) N_{\downarrow}(t)+N(t)+
\frac{\Gamma_{N}}{\pi}\int_{-\infty}^{\infty} d\varepsilon
\left[ f_{N}(\varepsilon)\Psi_{1}(\varepsilon,t)
+\Psi_{2}(\varepsilon,t)\right] ,
\label{A3}
\end{eqnarray}
where
\begin{eqnarray}
N_{\uparrow}(t) &=& \Gamma_{S}^{2}\lambda_{1}^{4}
\left( {\cal{L}}^{-1}\left\{ \frac{(s+\Gamma_{N})^{2}}{H_{3}(s)}\right\} (t)\right)^{2}
- \Gamma_{S}^{2} \left( {\cal{L}}^{-1}\left\{ \frac{H_{1}(s)}{H_{3}(s)}\right\} (t)\right)^{2} ,
\\
N_{\downarrow}(t) &=& 
\left( {\cal{L}}^{-1}\left\{ \frac{\Gamma_{S}^{2}H_{1}(s)}{(s+\Gamma_{N})H_{3}(s)} 
-\frac{1}{s+\Gamma_{N}}\right\} (t)\right)^{2} - \Gamma_{S}^{2} \lambda_{1}^{4}
\left( {\cal{L}}^{-1}\left\{ \frac{1}{H_{3}(s)}\right\} (t)\right)^{2} ,
\\
N(t) &=& \Gamma_{S}^{2}
\left( {\cal{L}}^{-1}\left\{ \frac{H_{1}(s)}{H_{3}(s)}\right\} (t)\right)^{2} +
\Gamma_{S}^{4}\lambda_{1}^{4} \left( {\cal{L}}^{-1}\left\{ \frac{1}{H_{3}(s)}\right\} (t)\right)^{2}
+ \frac{\Gamma_{S}^{2}\lambda_{1}^{2}}{2}
\left( {\cal{L}}^{-1}\left\{ \frac{1}{H_{2}(s)}\right\} (t)\right)^{2} ,
\end{eqnarray}
and
\begin{eqnarray}
\Psi_{1}(\varepsilon,t) &=& \Gamma_{S}^{2}\lambda_{1}^{4}
\left| {\cal{L}}^{-1}\left\{ \frac{s+\Gamma_{N}}{(s+i\varepsilon)H_{3}(s)}\right\} (t)\right|^{2}
- \Gamma_{S}^{2}\lambda_{1}^{4} 
\left| {\cal{L}}^{-1}\left\{ \frac{1}{(s+i\varepsilon)H_{3}(s)}\right\} (t)\right|^{2}
\\ &+&
\left| {\cal{L}}^{-1}\left\{ \left( \frac{\Gamma_{S}^{2}H_{1}(s)}{(s+\Gamma_{N})H_{3}(s)} 
-\frac{1}{s+\Gamma_{N}} \right) \frac{1}{s-i\varepsilon}\right\} (t)\right|^{2} - \Gamma_{S}^{2}
\left| {\cal{L}}^{-1}\left\{ \frac{H_{1}(s)}{(s-i\varepsilon)H_{3}(s)}\right\} (t)\right|^{2}
\\
\Psi_{2}(\varepsilon,t) &=& \Gamma_{S}^{4}\lambda_{1}^{4}
\left| {\cal{L}}^{-1}\left\{ \frac{1}{(s+i\varepsilon)H_{3}(s)} 
\right\} (t)\right|^{2} + \Gamma_{S}^{2} 
\left| {\cal{L}}^{-1}\left\{ \frac{H_{1}(s)}{(s+i\varepsilon)H_{3}(s)}\right\} (t)\right|^{2}.
\end{eqnarray}
Note, that $n_{1\downarrow}(t)$ does not depend on $\lambda_{2}$ and $\varepsilon_{2}$.

The intra-dor pairing function (\ref{intra-dot}) can be computed, using the Laplace
transforms $\hat{d}_{1\uparrow}(s)$ and $\hat{d}_{1\downarrow}(s)$ expressed 
in equations (\ref{d1up},\ref{d1down}). Following the precudure discussed above
for the charge occupancies we finally obtain (assuming the initial empty dots)
\begin{eqnarray}
\expect{ \hat{d}_{1\downarrow}(t) \hat{d}_{1\uparrow}(t) }
= D(t) + i\frac{\Gamma_{N}\Gamma_{S}}{\pi} \int_{-\infty}^{\infty}
\left[ f_{N}(\varepsilon) D_{1}(\varepsilon,t)+D_{2}(\varepsilon,t)\right] ,
\end{eqnarray}
where
\begin{eqnarray}
D(t) &=& i\Gamma_{S} \left[ \lambda_{1}^{4} {\cal{L}}^{-1}\left\{ \frac{s+\Gamma_{N}}
{H_{3}(s)} \right\} (t) {\cal{L}}^{-1}\left\{ \frac{(s+\Gamma_{N})^{2}}
{H_{3}(s)} \right\} (t)
+ {\cal{L}}^{-1}\left\{ \frac{H_{1}(s)}{H_{3}(s)} \right\} (t) 
{\cal{L}}^{-1}\left\{ \frac{\Gamma_{S}^{2}H_{1}(s)-H_{3}(s)}{(s+\Gamma_{N})H_{3}(s)} \right\} (t)
\right. \nonumber \\
&+& \left. \frac{\lambda_{1}^{2}}{2} {\cal{L}}^{-1}\left\{ \frac{1}{H_{2}(s)} \right\} (t) 
{\cal{L}}^{-1}\left\{ \frac{s+\Gamma_{N}}{H_{2}(s)} \right\} (t) \right] ,
\label{A11}
\\
D_{1}(\varepsilon,t) &=& \lambda_{1}^{4} {\cal{L}}^{-1}\left\{ \frac{s+\Gamma_{N}}
{(s+i\varepsilon)H_{3}(s)} \right\} (t) {\cal{L}}^{-1}\left\{ \frac{\Gamma_{S}^{2}-(s+\Gamma_{N})^{2}}
{(s-i\varepsilon)H_{3}(s)} \right\} (t)
\nonumber \\
&+& {\cal{L}}^{-1}\left\{ \frac{H_{1}(s)}{(s-i\varepsilon)H_{3}(s)} \right\} (t) 
{\cal{L}}^{-1}\left\{ \frac{(s+\Gamma_{N})^{2}H_{1}(s)-\Gamma_{S}^{2}H_{1}(s)+H_{3}(s)}
{(s+\Gamma_{N})(s+i\varepsilon)H_{3}(s)} \right\} (t) ,
\label{A12}
\\
D_{2}(\varepsilon,t) &=& \lambda_{1}^{4} {\cal{L}}^{-1}\left\{ \frac{s+\Gamma_{N}}
{(s+i\varepsilon)H_{3}(s)} \right\} (t) {\cal{L}}^{-1}\left\{ \frac{(s+\Gamma_{N})^{2}}
{(s-i\varepsilon)H_{3}(s)} \right\} (t)
\nonumber \\
&+& {\cal{L}}^{-1}\left\{ \frac{H_{1}(s)}{(s-i\varepsilon)H_{3}(s)} \right\} (t) 
{\cal{L}}^{-1}\left\{ \frac{\Gamma_{S}^{2}H_{1}(s)-H_{3}(s)}
{(s+\Gamma_{N})(s+i\varepsilon)H_{3}(s)} \right\} (t) .
\label{A13}
\end{eqnarray}
%


\twocolumngrid

\section{Transition probabilities for $\Gamma_{N}=0$ case}
\label{app_B}

In Sec.~\ref{charge_of_double_dots} we have shown that electron occupancy 
of QD$_{1}$ does not depend on the topological nanowire coupling to the second 
quantum dot. With this conclusion in mind, let us first consider a simplified
version of our setup, $\lambda_{2}=0$, in order to determine the charge occupancy 
of the proximitized QD$_1$ side-attached to the Majorana nanowire (i.e.\ 
completely ignoring any influence of QD$_{2}$). We choose the basis states
$\left| n_{1\uparrow},n_{1\downarrow},n_{f}\right>$, where $n_{1\sigma}$
represents either the empty or occupied $\sigma$-spin of QD$_{1}$ and $n_{f}$
stands for the number of nonlocal fermion constructed from the Majorana quasiparticles.

For specific considerations, we assume the initial ($t\leq 0$) configuration  
to be empty $\left| 0,0,0 \right>$. In what follows, we compute the time-dependent 
fillings after connecting the proximitized QD$_{1}$ to the topological nanowire. 
Expressing the time-dependent state vector by
\begin{eqnarray}
\left| n_{1\uparrow}(t),n_{1\downarrow}(t),n_{f}(t) \right> =
a_{1}(t) \left| 0,0,0 \right> + a_{2}(t) \left| 1,1,0 \right> 
\nonumber \\
+a_{3}(t) \left| 0,1,1 \right> + a_{4}(t) \left| 1,0,1 \right> 
\nonumber
\end{eqnarray}
we solve the Schr\"odinger equation to find the probability coefficients 
$a_{j}(t)$ for $t>0$. From straightforward calculations we obtain the following coefficients
\begin{eqnarray}
a_{1}(t) &=& \frac{1}{2} \left[ \left( \frac{\Gamma_{S}}{\sqrt{\Gamma_{S}^{2}
+2\lambda^{2}}}+1\right) \cos{\left(t\sqrt{b}\right)} \right.
\nonumber \\ 
&& - \left. \left( \frac{\Gamma_{S}}{\sqrt{\Gamma_{S}^{2}+2\lambda^{2}}}-1\right) 
\cos{\left(t\sqrt{c}\right)} \right] ,
\label{eqn_B1} \\
a_{2}(t) &=& \frac{i}{\sqrt{\Gamma_{S}^{2}+2\lambda^{2}}} \left[ \sqrt{c}
\sin{\left(t\sqrt{c}\right)} - \sqrt{b}\sin{\left(t\sqrt{b}\right)}  \right] ,
\label{eqn_B2} \\
a_{3}(t) &=& \frac{\lambda}{\sqrt{2\Gamma_{S}^{2}+(2\lambda)^{2}}}
\left[ \cos{\left(t\sqrt{c}\right)} - \cos{\left(t\sqrt{b}\right)}  \right] ,
\label{eqn_B3} \\
a_{4}(t) &=& \frac{i\lambda\sqrt{2}}{4} \left[ \left( \frac{\Gamma_{S}}{\sqrt{\Gamma_{S}^{2}
+2\lambda^{2}}}-1\right) \frac{\sin{\left(t\sqrt{c}\right)}}{\sqrt{c}} \right.
\nonumber \\ 
&& - \left. \left( \frac{\Gamma_{S}}{\sqrt{\Gamma_{S}^{2}+2\lambda^{2}}}+1\right) 
\frac{\sin{\left(t\sqrt{b}\right)}}{\sqrt{b}} \right] ,
\label{eqn_B4} 
\end{eqnarray}
where
\begin{eqnarray}
b/c= \frac{1}{2}\left( \lambda^{2}+\Gamma_{S}^{2}\pm \Gamma_{S}
\sqrt{\Gamma_{S}^{2}+2\lambda^{2}} \right) . \nonumber
\end{eqnarray}
The time-dependent occupancies can be expressed in terms of these  
coefficients as
\begin{eqnarray}
n_{1\uparrow}(t) &=& \left| a_{2}(t) \right|^{2} + \left| a_{4}(t) \right|^{2} ,
\label{eqn_B5} \\
n_{1\downarrow}(t) &=& \left| a_{2}(t) \right|^{2} + \left| a_{3}(t) \right|^{2} ,
\label{eqn_B6} \\
n_{f}(t) &=& \left| a_{3}(t) \right|^{2} + \left| a_{4}(t) \right|^{2}  
\label{eqn_B7}
\end{eqnarray}
and they are consistent with equations (\ref{eqn17},\ref{eqn18}) obtained 
in the main part for the $\Gamma_{N}=0$ case.

\begin{figure}[t]
\includegraphics[width=1\columnwidth]{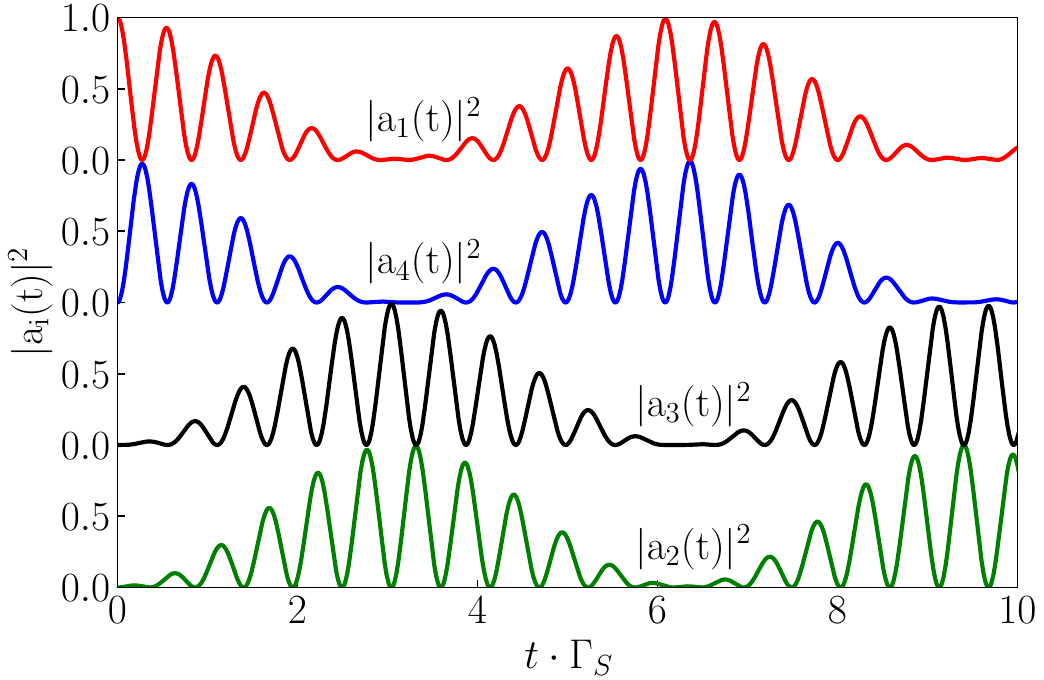}
\caption{Time-dependent probabilities $|a_{j}(t)|^{2}$ of a transition
from the initial empty configuration to the state $\left| j \right>$
defined in the basis $\left| n_{1\uparrow},n_{1\downarrow},n_{f} \right>$
obtained for the limit of $\Gamma_{N}=0$, $\lambda_{1}=8$,  $\lambda_{2}=0$.}
\label{Fig10}
\end{figure}

We can say that the formulas (\ref{eqn_B5}-\ref{eqn_B7}) resemble the Rabi 
oscillations of a four-level quantum system. In comparison with a two-level
quantum system (realized when the uncorrelated quantum dot is coupled to superconducting
lead in the limit of infinite pairing gap \cite{Taranko-2018}) we observe here the oscillation of the state vector between four quantum states.
We can describe the system evolution as an alternate oscillations between
$\left| 0,0,0 \right>$ and $\left| 1,0,1 \right>$ (two upper curves in
Fig.\ \ref{Fig10}) and between $\left| 0,1,1 \right>$ and $\left| 1,1,0 \right>$ states (two lower curves in Fig.\ \ref{Fig10}), respectively.

%

\begin{figure}[t]
\includegraphics[width=1\columnwidth]{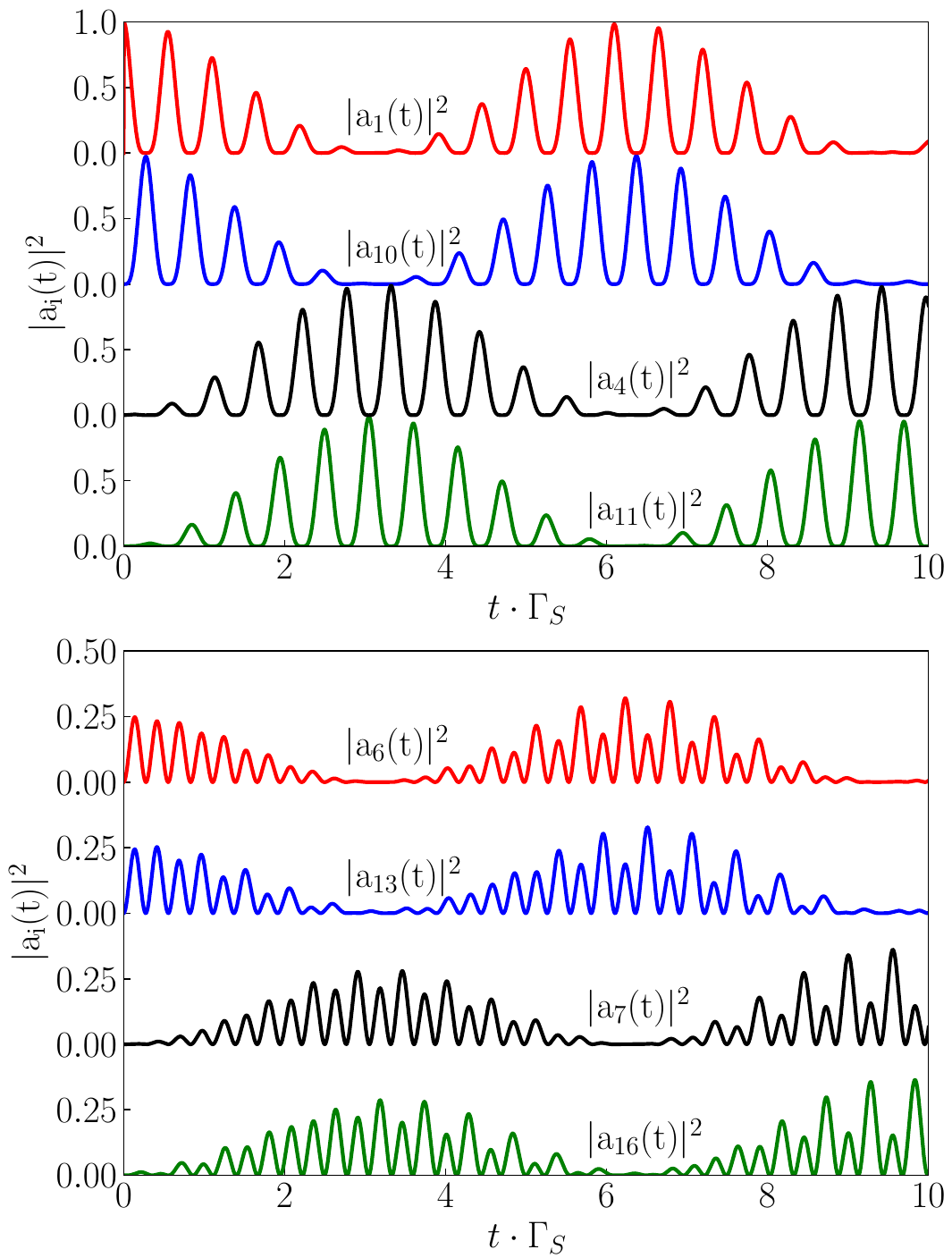}
\caption{Time-dependent probabilities $|a_{j}(t)|^{2}$ of a transition
from the initial empty configuration to the state $\left| j \right>$
defined in the basis $\left| n_{1\uparrow},n_{1\downarrow},n_{f},n_{2\uparrow} \right>$
obtained for $\Gamma_{N}=0$,  $\lambda_{1}=\lambda_{2}=8$.}
\label{Fig11}
\end{figure}

Note, that for the complete setup with both quantum dots the situation 
is more complicated because transitions would occur in larger basis
$\left| n_{1\uparrow},n_{1\downarrow},n_{f},n_{2\uparrow} \right>$
comprising 16 possible configurations. Eight states correspond to
even-parity and the other eight to odd-parity sectors. Assuming
the initially empty state $\left| 0,0,0,0 \right>$ we can express
the latter state vector by a linear combination of the eight even-parity
states with the corresponding time-dependent coefficients.

Introducing the auxiliary notation $\left|1\right>=\left|0,0,0,0 \right>$,
$\left|2\right>=\left|1,0,0,0 \right>$, $\left|3\right>=\left|0,1,0,0 \right>$,
$\left|4\right>=\left|1,1,0,0 \right>$, $\left|5\right>=\left|0,0,1,0 \right>$,
$\left|6\right>=\left|1,0,1,0 \right>$, $\left|7\right>=\left|0,1,1,0 \right>$,
$\left|8\right>=\left|1,1,1,0 \right>$, $\left|9\right>=\left|0,0,0,1 \right>$,
$\left|10\right>=\left|1,0,0,1 \right>$, $\left|11\right>=\left|0,1,0,1 \right>$,
$\left|12\right>=\left|1,1,0,1 \right>$, $\left|13\right>=\left|0,0,1,1 \right>$,
$\left|14\right>=\left|1,0,1,1 \right>$, $\left|15\right>=\left|0,1,1,1 \right>$,
$\left|16\right>=\left|1,1,1,1 \right>$, we can express the time-dependent
occupancies of QD$_{1}$ as
\begin{eqnarray}
n_{1\uparrow}(t) &=& \left| a_{4}(t) \right|^{2} + \left| a_{6}(t) \right|^{2}
+\left| a_{10}(t) \right|^{2} + \left| a_{16}(t) \right|^{2} ,
\label{eqn_B8} \\
n_{1\downarrow}(t) &=& \left| a_{4}(t) \right|^{2} + \left| a_{7}(t) \right|^{2}
+\left| a_{11}(t) \right|^{2} + \left| a_{16}(t) \right|^{2} .
\label{eqn_B9}
\end{eqnarray}
The spin-$\uparrow$ (spin-$\downarrow$) occupancy of QD$_{1}$ indicates 
that at a given instant of time the system can be found in configurations 
with the occupied second quantum dot $\left|1,0,0,1 \right>$ and 
$\left|1,1,1,1 \right>$ ($\left|0,1,0,1 \right>$ and $\left|1,1,1,1 \right>$).

In Fig.~\ref{Fig11} we plot the probabilities $|a_{j}(t)|^{2}$
for the corresponding states $\left| j \right>$, as indicated.
During the time evolution we observe clear oscillations between
$\left|0,0,0,0 \right>$ and $\left|1,0,0,1 \right>$ states
(two upper curves in upper panel in Fig. \ref{Fig11}) and simultaneously
oscillations between the states $\left|1,1,0,0 \right>$ and 
$\left|0,1,0,1 \right>$ (two lower curves in upper panel in Fig. \ref{Fig11}).


From such considerations, we can also determine the pairing functions,
expressing them by the complex coefficients $a_{j}(t)$. For the initial 
empty configuration, the on-dot pairing (\ref{intra-dot}) takes the
following form
\begin{equation}
C_{11}(t)=a_{1}^{\star}(t)a_{4}(t) + a_{13}^{\star}(t) a_{16}(t) ,
\label{eqn_B10}
\end{equation}
whereas for the initial odd-parity one obtains 
\begin{equation}
C_{11}(t)=a_{9}^{\star}(t)a_{12}(t) + a_{5}^{\star}(t) a_{8}(t) .
\label{eqn_B11}
\end{equation}
In other words, the on-dot pairing function depends on the amplitude probabilities that the system evolving  over all basis states can be found in states  $\left|0,0,0,0 \right>$, 
$\left|1,1,0,0 \right>$, $\left|0,0,1,1 \right>$ and $\left|1,1,1,1 \right>$,
respectively. 

Similarly, we can determine the non-local pairing functions.
For the initial empty configurations they are given by
\begin{eqnarray}
C_{12}(t) &=& a_{6}^{\star}(t)a_{16}(t) - a_{1}^{\star}(t) a_{11}(t),
\label{eqn_B12} \\
C_{2f}(t) &=& a_{4}^{\star}(t)a_{16}(t) + a_{1}^{\star}(t) a_{13}(t) .
\label{eqn_B13}
\end{eqnarray}
Note, that (\ref{eqn_B10},\ref{eqn_B13}) depend on the same coefficients
$a_{j}(t)$, but in different combinations.

\section{Nonlocal triplet pairing}
\label{nonlocal_triplet}

\begin{figure}[b]
\includegraphics[width=0.95\columnwidth]{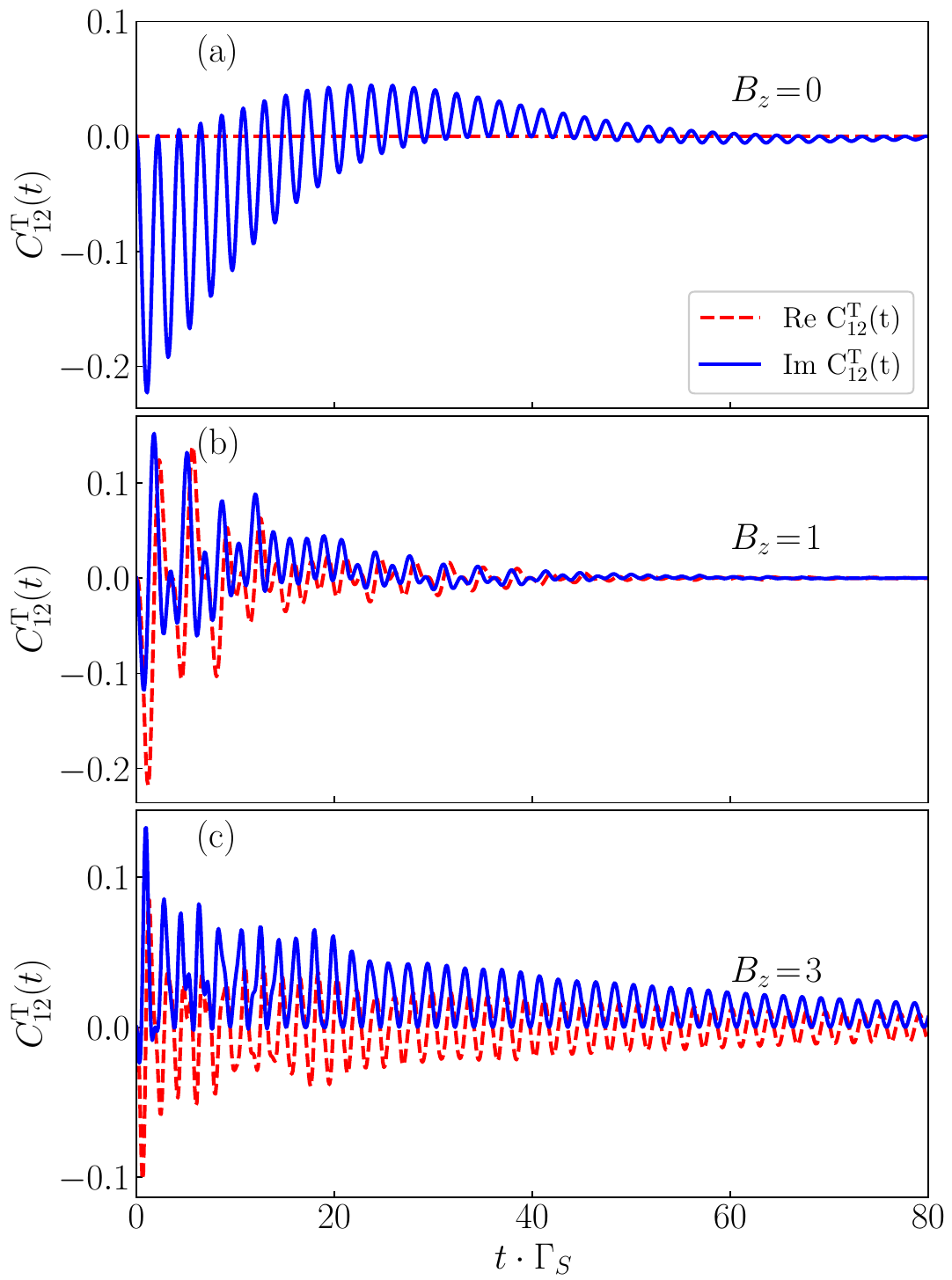}
\caption{
{Transient dynamics of the nonlocal pairing $C_{12}^{\rm T}(t)$ 
in the triplet channel obtained for several values of the Zeeman field $B_{z}$, using the same set of model parameters 
as in Fig.~\ref{Fig_new_1}.}}
\label{Fig_new_3}
\end{figure}

\begin{figure}
\includegraphics[width=0.95\columnwidth]{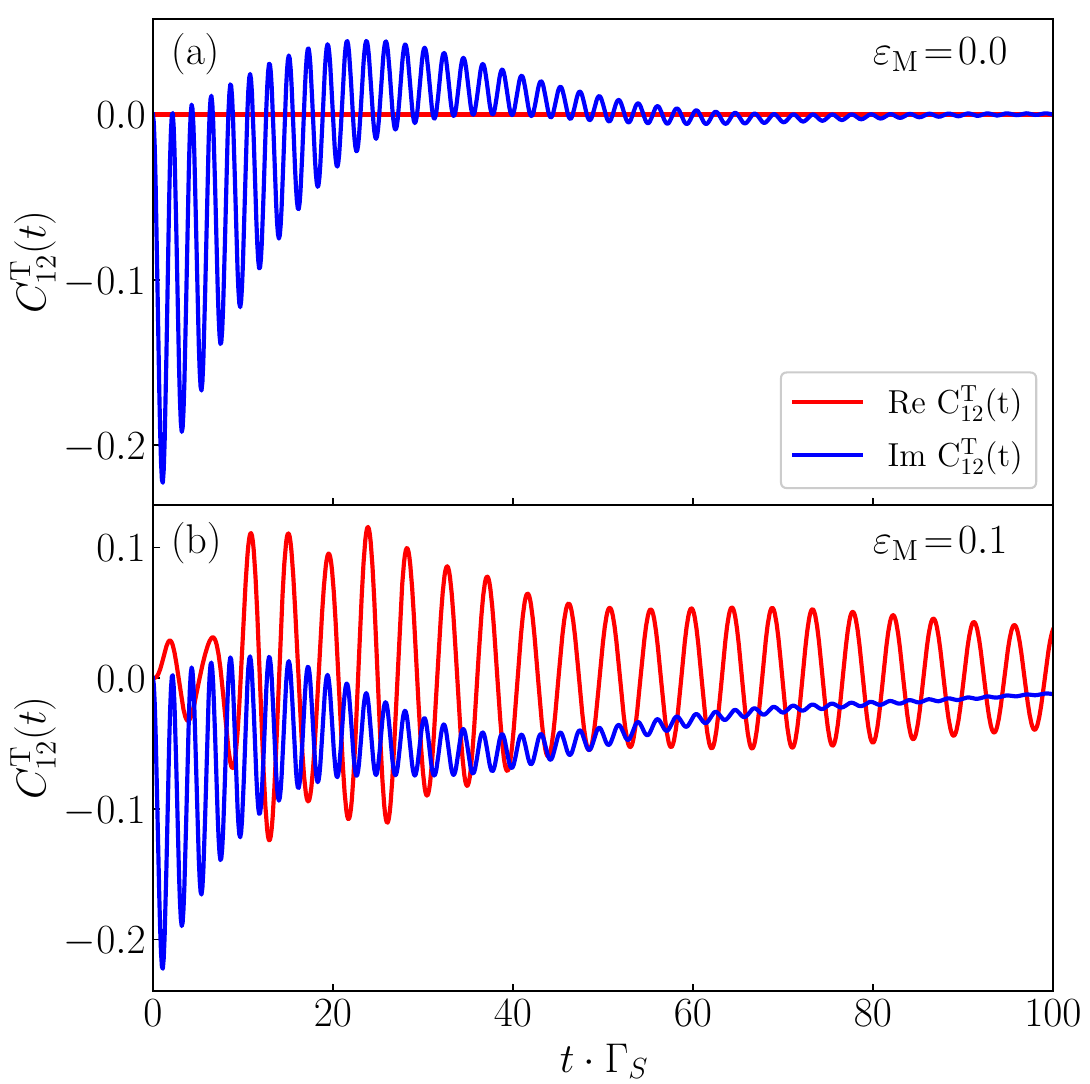}
\caption{
{Transient dynamics of the nonlocal pairing $C_{12}^{\rm T}(t)$ 
in the triplet channel obtained for the nonoverlaping (upper panel) 
and overlaping Majorana modes (bottom panel), 
using the same model parameters as in Fig.~\ref{Fig_new_1}.}}
\label{Fig_new_4}
\end{figure}

The leakage of Majorana modes onto the side-attached quantum dots is
strictly related to the intersite triplet pairing between the outer
sites of the topological nanowire and quantum dots. Within the present
low-energy approach, such mechanism is captured by the mixed pairing
$\expect{ \hat{d}_{i\uparrow}(t)\hat{f}(t)}$, as discussed in 
Sec.~\ref{inter_site_triplet}. In this context, it is natural to explore
the possible emergence of the nonlocal triplet pairing
\begin{eqnarray}
C_{12}^{\rm T}(t)= \expect{ \hat{d}_{1\uparrow}(t)\hat{d}_{2\uparrow}(t)} 
\label{triplet_def}
\end{eqnarray}
because its efficiency might be detectable using the spin-polarized 
crossed Andreev reflection spectroscopy. Adopting our methodology to the
nonlocal triplet pairing (\ref{triplet_def}), we obtain for $\varepsilon_{i\sigma}=0$ and $\epsilon_{M}=0$
\begin{eqnarray}
C_{12}^{\rm T}(t) &=& \lambda_{1}\lambda_{2}
\left(  n_{f}(0)-\frac{1}{2}\right) 
\nonumber \\ & \times &
{\cal{L}}^{-1}\left\{ \frac{s+g}{H_{2}{s}} \right\}(t)
{\cal{L}}^{-1}\left\{ \frac{1}{s^{2}+2\lambda_{2}^{2}} \right\}(t).
\end{eqnarray}
In particular,  for $\Gamma_{N}=0$, this formula simplifies to 
\begin{eqnarray}
C_{12}^{\rm T}(t) &=& \frac{\lambda_{1}}{\sqrt{2\Delta^{2}+4\lambda_{1}^{2}}}
\left(  n_{f}(0)-\frac{1}{2}\right) 
\nonumber \\ & \times &
\sin{\left( \sqrt{2}\lambda_{2}t \right)}
\sin{\left( \sqrt{\Delta^{2}+2\lambda_{1}^{2}}t \right)} ,
\end{eqnarray}
indicating that inter-dot triplet pairing occurs exclusively when both couplings
$\lambda_{1,2}\neq 0$. For vanishing $\Gamma_{N}$, the relaxation mechanism 
is blocked, hence under such conditions, $C_{12}^{\rm T}(t)$ acquires oscillatory behavior with a convolution of the frequency $\sqrt{2}\lambda_{2}$ and
$\sqrt{2}\lambda_{1}\sqrt{1+\frac{\Delta^{2}}{2\lambda_{1}^{2}}}$.

More general results obtained numerically for finite $\Gamma_{N}$ are presented in Figs. \ref{Fig_new_3} and \ref{Fig_new_4}. Again we notice that the triplet nonlocal
pairing survives only temporarily, in similar fashion as the singlet nonlocal pairing does. Magnetic field affects the profile of quantum oscillations, however, in contrast to the singlet inter-dot pairing, it seems that $C_{12}^{\rm T}(t)$ survives over pretty long time-scale upon increasing $B_{z}$ (Fig. \ref{Fig_new_3}). Apparently, this is related to the coexistence of magnetism and triplet pairing \cite{Buzdin_2005}. Hybridization of boundary modes, $\epsilon_{M}$, is another factor that prolongs the existence of nonlocal triplet pairing (Fig. \ref{Fig_new_4}). Such an effect is perhaps less surprising, because the degree of overlapping Majorana modes goes hand in hand with the shortening of the topological nanowire, which can be expected to favor the mutual inter-dot pairing.

\bibliography{myBib}


\end{document}